\documentclass[aps,prd,preprint,superscriptaddress]{revtex4-1}
\usepackage{epsfig,dsfont,amssymb,amsmath,amsthm,amsfonts,amsbsy,mathrsfs,mathtools,appendix}
\usepackage{graphicx}
\usepackage{float}
\usepackage{color}
\usepackage{cases}
\usepackage{scrtime}
\usepackage{rotating}
\usepackage{multirow} 
\usepackage{CJKutf8}
\usepackage{enumerate}

\newcommand\aap{Astron. Astrophys.}

\newcommand\apjl{Astrophys. J. Lett.}

\newcommand\cqg{Class. Quantum Gravity}

\newcommand\epjc{Eur. Phys. J. C}

\newcommand\grg{Gen. Relativ. Gravit.}

\newcommand\ijmpd{Int. J. Mod. Phys. D}

\newcommand\jcap{J. Cosmol. Astropart. Phys.}

\newcommand\lrr{Living Rev. Relativ.}
\newcommand\mnras{Mon. Not. R. Astron. Soc.}

\newcommand\natast{Nat. Astron.}

\newcommand\na{New Astron.}

\newcommand\physrep{Phys. Rep.}

\newcommand\plb{Phys. Lett. B}

\newcommand\pr{Phys. Rev.}
\newcommand\prx{Phys. Rev. X}

\newcommand\pnas{Proc. Natl. Acad. Sci. U.S.A.}

\newcommand\sciadv{Sci. Adv.}

\newcommand\SchStar{Schwarzschild star\ }

\begin{document}
\title{Higher-order inner photon rings of a horizonless ultracompact object with an antiphoton sphere and their interferometric pattern}

\author{Yuan-Xing Gao}
\email{gao$_$yx@outlook.com}
\affiliation{Purple Mountain Observatory, Chinese Academy of Sciences, Nanjing 210023, China
}

\begin{abstract}
A horizonless ultracompact object can have a stable antiphoton sphere, which causes the strong deflection of photons inside the unstable photon sphere, leading to the formation of distinctive inner photon rings.
In this work, we present analytical descriptions for the shape, thickness and interference pattern of higher-order inner photon rings.
By taking the static spherically symmetric Schwarzschild star with a photon sphere as an example, we find that its inner photon rings can be more non-circular and thicker than the outer ones, and show that the inclusion of  the inner photon rings can give rise to new features in the interferometric pattern.
Our formulae can also be applied to other ultracompact objects, providing a convenient way to study the observational properties of their higher-order photon rings.
\end{abstract}

\maketitle

\allowdisplaybreaks


\section{Introduction}
\label{sec:intro}

A black hole is the fundamental object in general relativity, its abundance in the Universe has been demonstrated observationally, such as the detection of gravitational waves from merging binary black holes \cite{LVC2016PRL116.061102, LVC2016PRX6.041015, LVC2016PRL116.241103, LVC2017PRL118.221101, LVC2017ApJ851.L35, LVC2017PRL119.141101} and direct images of the supermassive black holes M87* \cite{EHTC2019ApJ875.L1, EHTC2019ApJ875.L2, EHTC2019ApJ875.L3, EHTC2019ApJ875.L4, EHTC2019ApJ875.L5, EHTC2019ApJ875.L6} and Sgr A* \cite{EHTC2022ApJ930.L12,EHTC2022ApJ930.L13,EHTC2022ApJ930.L14,EHTC2022ApJ930.L15,EHTC2022ApJ930.L16,EHTC2022ApJ930.L17}.
However, the black hole has an event horizon that blocks its connection between the interior and exterior regions, resulting in Hawking radiation and the information paradox \cite{Hawking1975CMP43.199,Mathur2009CQG26.224001}, and a singularity that straightforwardly causes the breakdown of general relativity.
In order to solve these problems, ultracompact objects that possess an unstable photon sphere but no event horizon have been put forth in the literature (see Ref.~\cite{Cardoso2019LRR22.4} for a review).
One prominent example is the horizonless Schwarzschild star with a photon sphere, which removes the intrinsic singularity and event horizon of the Schwarzschild metric by introducing a bounded isotropic fluid into a region with radius slightly larger than the Schwarzschild radius \cite{Schwarzschild1916SPAWB424,Wald1984Book}.

Due to the absence of the event horizon, an ultracompact object can generate unique echoes in the ringdown phase \cite{Cardoso2017NatAst1.586}. 
Therefore, gravitational wave observations provide a way to distinguish it from a black hole. 
However,  it is not until the next generation of space-borne gravitational wave detectors come into operation that we can separate these echoes from the waveforms \cite{Berti2016PRL117.101102}.

One can also identify an ultracompact object by electromagnetic observations. 
A part of the ultracompact object models  have been ruled out by the constraints from the size, shape, absorption and albedo of the shadow region in the image taken by the Event Horizon Telescope \cite{EHTC2019ApJ875.L5, EHTC2022ApJ930.L17}. 
To further explore the possibility of the survived ones, more sophisticated data processing is required. 
However, given that the measurement of the shadow itself strongly depends on the accretion disk model \cite{EHTC2019ApJ875.L5,EHTC2022ApJ930.L16} and the accretion physics in these observations is still poorly understood, it is challenging in current stage to obtain more precise results based on the observed shadow only.

Hidden in the image of a compact object are other subtle structures, such as the relativistic images and photon rings, which are potentially to be resolved by future space-borne very long baseline interferometry \cite{Gralla2019PRD100.024018,Johnson2020SciAdv6.eaaz1310,Vincent2022AAp667.A170,Paugnat2022AAp668.A11}.
Relativistic images are formed by photons that wind around the black hole or ultracompact object one or more times before reaching the observer. Their corresponding emitters are typically point-like discrete sources, such as hotspots and compact stars \cite{Virbhadra2000PRD62.084003,Bozza2001GRG33.1535,Bozza2007PRD76.083008,Shaikh2019PRD99.104040}. The electromagnetic signals of these images have been studied in detail \cite{Petters2003MNRAS338.457,Aratore2021JCAP10.054,Gao2022EPJC82.162,Gao2024PRD109.063030,Chen2024JCAP04.032}. Especially in interferometry, the observational signatures of relativistic images exhibit an oscillating staircase-like structure, revealing their detectability with an Earth-Moon baseline \cite{Aratore2021JCAP10.054,Gao2024PRD109.063030}. 
In comparison, photon rings arise from the lensed photons emitted by continuous extended sources like the accretion disk \cite{Gralla2019PRD100.024018,Johnson2020SciAdv6.eaaz1310,Vincent2022AAp667.A170,Paugnat2022AAp668.A11,Tsupko2022PRD106.064033,CardenasAvendano2023PRD108.064043,Aratore2024PRD109.124057,Jia2024PRD110.083044}. On the image plane of the observer, a photon ring presents a ring-shaped structure, hence the name.
Since photon rings are also generated by the strong deflection of photons around the black hole or ultracompact object, they can be regarded as a particular realization of relativistic images.
The order of a photon ring is commonly denoted by the number of half orbits $n$ \cite{Gralla2019PRD100.024018,Tsupko2022PRD106.064033}. 
Photon rings with large $n$ asymptotically approach the boundary of black hole shadow \cite{Falcke2000ApJ528.L13,Cunha2018GRG50.42,Perlick2022PR947.1,Paugnat2022AAp668.A11}.
The bright thick rings surrounding M87* and Sgr A* observed by the Event Horizon Telescope can be interpreted as the lowest-order $n=0$ and $n=1$ photon rings, which are highly dependent on the accretion disk model and accretion physics \cite{EHTC2019ApJ875.L1, EHTC2022ApJ930.L12}.
It is shown that ($n \geqslant 2$) higher-order photon rings could give a good approximation of the black hole shadow border, and their shape and size are mainly determined by the spacetime geometry \cite{Paugnat2022AAp668.A11}.  
Thus, those rings can be used to test the theory of gravity. 
Compared with current direct images of the accretion disk, the $n = 2$ photon ring is less dependent on the accretion model \cite{Vincent2022AAp667.A170}, reducing significantly the number of input parameters for data interpretation and thus being a better tool to investigate the properties of the central compact objects.

To date investigations on the photon rings are mostly numerical and computationally expensive \cite{Johnson2020SciAdv6.eaaz1310,Vincent2022AAp667.A170,Paugnat2022AAp668.A11,CardenasAvendano2023PRD108.064043}. 
In comparison, the strong deflection limit method for a black hole spacetime provides a cheaper way to study the properties of photon rings analytically \cite{Bozza2002PRD66.103001,Bozza2007PRD76.083008}, which has been successfully applied in the Schwarzschild and other black hole spacetimes \cite{Tsupko2022PRD106.064033,Aratore2024PRD109.124057}. This method is also extended to the ultracompact object spacetime \cite{Shaikh2019PRD99.104040},
and we make a further generalization of \cite{Shaikh2019PRD99.104040} by including the finite distance effect of a source \cite{Gao2024PRD109.063030}.
These results provide an analytical framework to investigate the difference between a black hole and an ultracompact object in the photon rings.

In addition to the unstable photon sphere, an ultracompact object can also have an antiphoton sphere, which defines a stable circular orbit in the static spherically symmetric case \cite{Patil2017PRD95.024026,Shaikh2019PRD99.104040,Gyulchev2021EPJC81.885,Zhu2020EPJC80.444,Gao2022EPJC82.162,Igata2025PRD111.084062}.
Owing to the presence of the antiphoton sphere, photons entering the photon sphere may be strongly deflected, and then be received by the observer and produce inner relativistic images \cite{Shaikh2019PRD99.104040,Tamm2024PRD109.044062,Zhu2020EPJC80.444,Gao2022EPJC82.162,Gao2024PRD109.063030,Chen2024JCAP04.032} or inner photon rings \cite{Olmo2021PLB829.137045,Gyulchev2021EPJC81.885,Rosa2023PRD107.084048}.
These unique features are generally absent in the black hole spacetime and are thus key to distinguish the ultracompact object from the black hole.
In our previous work \cite{Gao2024PRD109.063030}, we analytically studied the characteristics of inner relativistic images formed by point-like sources and estimated the detectability of these images. As a complement, this work focuses on the observational signatures of inner photon rings formed by photons emitted by the accretion disk, aiming to gain a batter understanding of the differences between ultracompact objects and black holes.

The shape of inner photon rings can deviate significantly from a circle and contribute moderately to the observed flux \cite{Gyulchev2021EPJC81.885}. Their existence might explain the radiation in the central dark region of the images detected by the Event Horizon Telescope \cite{EHTC2022ApJ930.L17}. However, due to the much smaller widths of inner photon rings compared to the primary disk and the limited interferometric baseline length, the current resolution of approximately 10 $\mu$as is still unable to resolve these rings, thereby leading to a lack of detailed studies on their interferometric signatures. To address this gap, as a preliminary study, we will give analytical descriptions for the shape, thickness and interferometric pattern of the higher-order photon rings around an ultracompact object based on the obtained results of the deflection angle by employing the strong deflection limit methods with the finite distance effect \cite{Bozza2007PRD76.083008,Gao2024PRD109.063030}.
Furthermore, this work establishes a connection between the analytical strong deflection limit method and the calculation of the interferometric visibility, providing a convenient way to explore the observational characteristics of the ultracompact objects in the radio wavelength band.

This paper is organized as follows. 
In Sec.~\ref{sec:chapter2}, we introduce the photon and antiphoton spheres that are of vital importance to the formation of the photon rings, and briefly review the strong deflection limit method for an ultracompact object.
Based on this method, 
in Sec.~\ref{sec:chapter3}, we analytically study the shape and thickness of the higher-order photon rings around an ultracompact object. 
In Sec.~\ref{sec:chapter4}, we further calculate the interferometric pattern of these rings analytically and compare it with the numerical one.
In Sec.~\ref{sec:chapter5}, by taking the horizonless Schwarzschild star with a photon sphere as a specific example, we study the shape, thickness and interferometric pattern of its $n=2$ inner and outer photon rings.
In Sec.~\ref{sec:conlude}, we conclude and discuss our results.

\section{Ultracompact object spacetime and strongly deflected photons}
\label{sec:chapter2}

The line element of a static spherically symmetric spacetime reads
\begin{equation}\label{MetricSchStar}
\mathrm{d}s^2=-A(r)\mathrm{d}t^2+B(r)\mathrm{d}r^2+C(r)(\mathrm{d}\theta^2+\sin^2\theta\mathrm{d}\phi^2),
\end{equation}
in which $A(r), B(r)$ and $C(r)$ are non-negative metric functions.
To describe a horizonless ultracompact object, the condition that the above metric has no event horizon but features a photon sphere is required.

In an ultracompact object spacetime, a photon may travel along a geodesic with the following equation of motion
\begin{eqnarray}
\label{GeodesicEquationOfSchStar}
A(r)B(r)\ \dot{r}^2 + L^2V_\mathrm{eff\bullet} = E^2,
\end{eqnarray}
where $E$ and $L$ are respectively the energy and angular momentum of the photon, 
and they are related to the impact parameter $u$ with \cite{Weinberg1972Book,Virbhadra1998AA337.1}
\begin{equation}
u\equiv\frac{L}{E}=\sqrt{\frac{C(r_0)}{A(r_0)}}.
\end{equation}
Here $r_0$ is the turning point of the photon.
The effective potential per $L^2$ is defined as \cite{Shaikh2019PRD99.104040}
\begin{equation}
    V_\mathrm{eff\bullet}=\frac{A(r)}{C(r)}.
\end{equation}
It is demonstrated that an ultracompact object always have two circular orbits, one unstable and the other stable \cite{Cunha2017PRL119.251102, Guo2020PRD103.104031}, which respectively correspond to the maximum and minimum of $V_\mathrm{eff\bullet}$ \cite{Shaikh2019PRD99.104040}.

\begin{figure}[H]    
	\centering
	\includegraphics[width=1\textwidth]{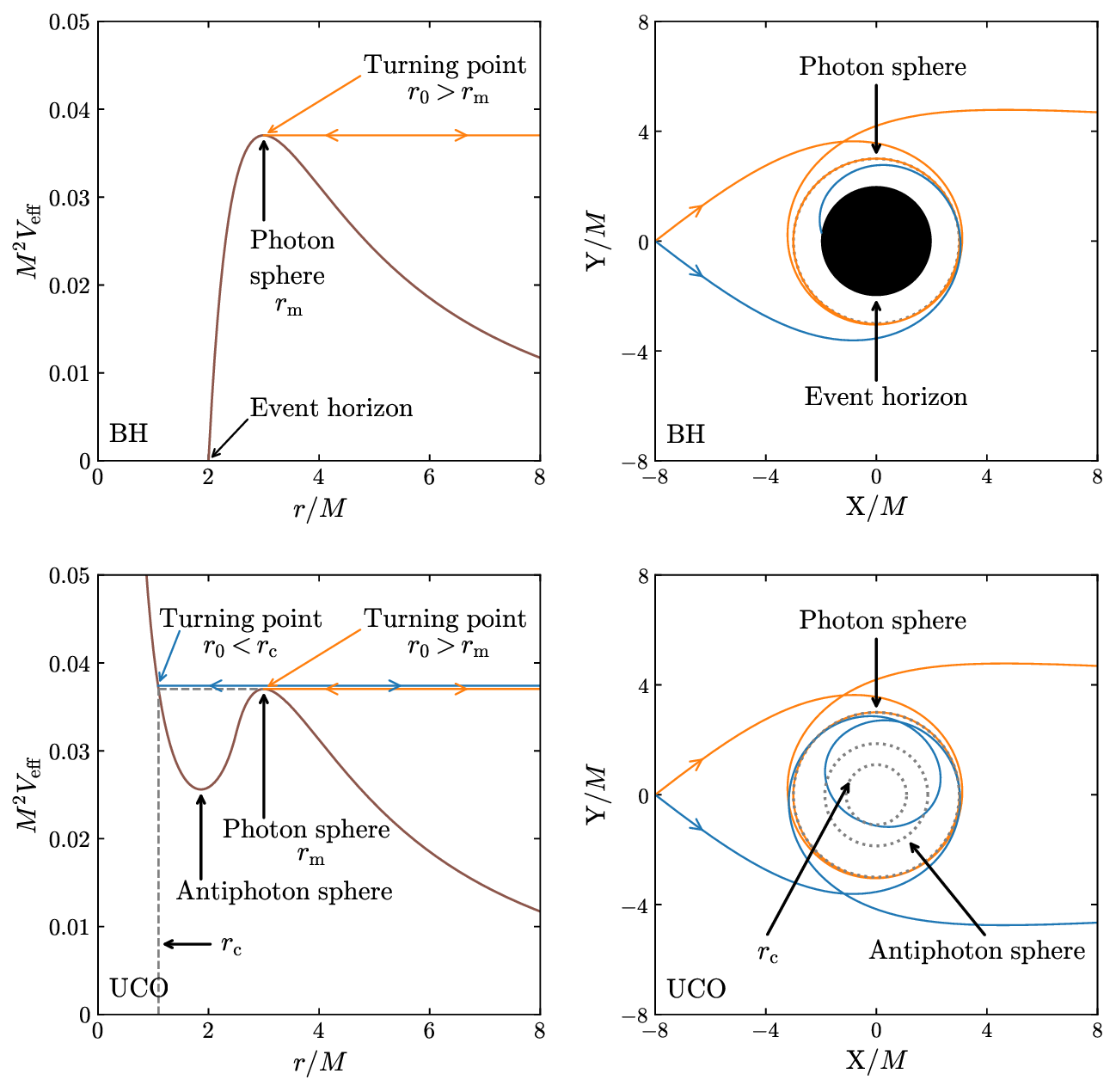}
	\caption{The effective potential $V_\mathrm{eff}$ and photon trajectories in the black hole spacetime (first row) and in the ultracompact object spacetime (second row). Both the black hole and the ultracompact object have a photon sphere located at $r = r_\mathrm{m}$, thus the deflection of photons outside their photon sphere is similar (orange line). Unlike the black hole,  the ultracompact object does not have an event horizon but has a unique antiphoton sphere and a spherical surface $r = r_\mathrm{c}$ with $V_\mathrm{eff} ( r_\mathrm{c} ) = V_\mathrm{eff} ( r_\mathrm{m} )$, causing photons entering the ultracompact object's photon sphere to escape (blue line). Here we use the Schwarzschild black hole (BH) and Schwarzschild star (UCO) to illustrate and set $G=c=1$.}
	\label{FigSup:Veff_rm_ra_rc}
\end{figure}

$\mathrm{d}V_\mathrm{eff\bullet}/\mathrm{d}r=0$ and $\mathrm{d}^2V_\mathrm{eff\bullet}/\mathrm{d}r^2<0$ define the unstable circular orbit, i.e., the photon sphere \cite{Bozza2002PRD66.103001}. 
We denote its radius by $r_\mathrm{m}$.
As shown in Fig.~\ref{FigSup:Veff_rm_ra_rc}, the photon with $r_0 \to r_\mathrm{m}^+$ (i.e., $r_0 > r_\mathrm{m}$ and $r_0 \to r_\mathrm{m}$) will be strongly deflected and wind around the photon sphere several times before it is received by the observer (orange line), while the photon that enters the photon sphere of a black hole will be absorbed by the event horizon (blue line in the first row) \cite{Bozza2002PRD66.103001}.

$\mathrm{d}V_\mathrm{eff\bullet}/\mathrm{d}r=0$ and $\mathrm{d}^2V_\mathrm{eff\bullet}/\mathrm{d}r^2>0$  define a stable circular orbit, i.e., the antiphoton sphere \cite{Shaikh2019PRD99.104040}. 
In general, the stable antiphoton sphere is a unique character of the ultracompact object, though there are few cases of black holes with an antiphoton sphere \cite{Cunha2016PRD94.104023}. The radius of the antiphoton sphere is smaller than that of the photon sphere, and it is a key factor for imaging inside the photon sphere \cite{Patil2017PRD95.024026,Shaikh2019PRD99.104040,Zhu2020EPJC80.444,Gao2022EPJC82.162,Igata2025PRD111.084062}.
Due to the existence of the antiphoton sphere, there is a radius $r_\mathrm{c}$ ($< r_\mathrm{m}$) inside the photon sphere, which satisfies 
\begin{eqnarray}\label{rcdef}
\frac{A(r_\mathrm{c})}{C(r_\mathrm{c})}=\frac{A(r_\mathrm{m})}{C(r_\mathrm{m})},
\end{eqnarray}
or $u(r_\mathrm{c})=u_\mathrm{m}$.
Note that $r_\mathrm{c}$ is not the radius of the antiphoton sphere, but it is always smaller than the latter.
As shown by the blue line in Fig.~\ref{FigSup:Veff_rm_ra_rc}, the presence of $r_\mathrm{c}$ leads to the fact that a photon with impact parameter less than $u_\mathrm{m}$ can have a turning point $r_0 < r_\mathrm{m}$. That photon may escape from the interior, as long as the surface of the ultracompact object does not absorb it. 
In the limit $r_0\to r_\mathrm{c}^-$ (i.e., $r_0 < r_\mathrm{c}$ and $r_0 \to r_\mathrm{c}$), a photon can also be strongly deflected \cite{Shaikh2019PRD99.104040}. An example is illustrated by the blue line in the second row of Fig.~\ref{FigSup:Veff_rm_ra_rc}. It should be emphasized that although it is true that the presence of an antiphoton sphere allows light rays to penetrate within the photon sphere, the dominant contribution to the total deflection still occurs near the photon sphere.

The strong deflection of photons in the ultracompact object spacetime is related to the change in the azimuthal angle $\Delta\phi$. When the spacetime is asymptotically flat, $\Delta\phi$ is defined as \cite{Bozza2007PRD76.083008}
\begin{eqnarray}\label{DeltaPhiInSchStarST}
\Delta\phi &=&\sum_{\mathrm{i}=\mathrm{S,O}}\int_{r_0}^{r_\mathrm{i}}\sqrt{\frac{C_0A(r)B(r)}{C(r)[A_0C(r)-A(r)C_0]}}\ \mathrm{d}r,
\end{eqnarray}
where the subscript $``0"$ denotes functions evaluated at $r=r_0$. $r_\mathrm{S}$ and $r_\mathrm{O}$ represent the radial coordinates of the source and the observer, respectively. 
In this work, we consider that the photon is emitted by a source located outside the photon sphere (i.e., a thin circular emission disk) with an initially radially-inward direction and can finally reach asymptotic observers \cite{Gao2024PRD109.063030}, thus we have $r_\mathrm{S} > r_\mathrm{m}$ and  $r_\mathrm{O} \gg r_\mathrm{m}$.
In the strong deflection limit $r_0\to r_\mathrm{m}^+$ and $r_0\to r_\mathrm{c}^-$, $\Delta\phi$ will be much larger than 1 and can be solved analytically \cite{Bozza2002PRD66.103001, Bozza2007PRD76.083008, Shaikh2019PRD99.104040, Gao2024PRD109.063030}.

\subsection{Strong deflection of photons outside the photon sphere}
In the limit of $r_0\to r_\mathrm{m}^+$, a photon with impact parameter $u\to u_\mathrm{m}^+$ will wind around the gravitational lens several times before being received by the observer.
Up to $\mathcal{O}[(u-u_\mathrm{m})\log(u-u_\mathrm{m})]$, the change in the azimuthal angle of that photon can be analytically obtained as \cite{Bozza2007PRD76.083008,Aratore2021JCAP10.054}
\begin{equation}\label{Delta_phi_rm_out}
\Delta\phi=-\bar{a}_+\log\frac{\epsilon_+}{z_\mathrm{O}z_\mathrm{S}}+\bar{b}_++\pi,
\end{equation}
where
\begin{eqnarray}
  \label{epsilon_p}
  \epsilon_+ & = & \frac{u}{u_\mathrm{m}}-1, \\
  \label{zdef}
  z_\mathrm{O} & = & 1-\frac{r_\mathrm{m}}{r_\mathrm{O}}, \ \ z_\mathrm{S}=1-\frac{r_\mathrm{m}}{r_\mathrm{S}}, \ \ z=1-\frac{r_\mathrm{m}}{r},\\
  \label{}
  \bar{a}_+ & = & \sqrt{\frac{2B_\mathrm{m}A_\mathrm{m}}{C_\mathrm{m}''A_\mathrm{m}-C_\mathrm{m}A''_\mathrm{m}}},\\
  \label{}
  \bar{b}_+ & = & -\pi+\bar{a}_+\log\bigg[r_\mathrm{m}^2\bigg(\frac{C_\mathrm{m}''}{C_\mathrm{m}}-\frac{A_\mathrm{m}''}{A_\mathrm{m}}\bigg)\bigg]\nonumber\\
  && +\bigg(\int^{z_\mathrm{O}}_{0}+\int^{z_\mathrm{S}}_{0}\bigg)g_1(z)\mathrm{d}z,\\
  \label{}
  g_1(z) & = & \frac{r_\mathrm{m}}{(1-z)^2}\sqrt{\frac{C_\mathrm{m}A(z)B(z)}{[A_\mathrm{m}C(z)-A(z)C_\mathrm{m}]C(z)}} \nonumber\\
  && -\frac{\bar{a}_+}{|z|},
\end{eqnarray}
The subscript $``\mathrm{m}"$ denotes quantities evaluated at $r=r_\mathrm{m}$.
From Eq.~\eqref{Delta_phi_rm_out}, we can obtain the impact parameter of the strongly deflected photon outside the photon sphere as
\begin{equation}\label{u_PS_out}
u_+ = u_\mathrm{m}\left(1+z_\mathrm{O}z_\mathrm{S}e^{\frac{\bar{b}_++\pi-\Delta\phi}{\bar{a}_+}}\right).
\end{equation}

\subsection{Strong deflection of photons inside the photon sphere}

In the limit $r_0\to r_\mathrm{c}^-$, 
a photon with $u\to u_\mathrm{m}^-$ can enter the region within the photon sphere. 
Due to the presence of the antiphoton sphere,
it is possible for the photon that enters the photon sphere to reach the observer after winding around the ultracompact object several times \cite{Shaikh2019PRD99.104040}.
Up to $\mathcal{O}[(u_\mathrm{m}^2-u^2)\log(u_\mathrm{m}^2-u^2)]$, the change in the azimuthal angle can be found as \cite{Gao2024PRD109.063030}
\begin{eqnarray}\label{Delta_phi_rm_in}
\Delta\phi = -\bar{a}_-\log\frac{\epsilon_-}{\sqrt{z_\mathrm{O}z_\mathrm{S}}}+\bar{b}_-+\pi,
\end{eqnarray}
where
\begin{eqnarray}
  \label{epsilon_m}
  \epsilon_- & = & \frac{u_\mathrm{m}^2}{u^2}-1, \\
  \label{abar_m}
  \bar{a}_- & = & 2\bar{a}_+,\\
  \label{}
  \bar{b}_- & = & -\pi+2\bar{a}_+\log\bigg[2r_\mathrm{m}^2\bigg(\frac{C_\mathrm{m}''}{C_\mathrm{m}}-\frac{A_\mathrm{m}''}{A_\mathrm{m}}\bigg)\bigg(\frac{r_\mathrm{m}}{r_\mathrm{c}}-1\bigg)\bigg] \nonumber\\
  & &+\bigg(\int^{z_\mathrm{O}}_{1-\frac{r_\mathrm{m}}{r_\mathrm{c}}}+\int^{z_\mathrm{S}}_{1-\frac{r_\mathrm{m}}{r_\mathrm{c}}}\bigg)g_1(z)\mathrm{d}z.
\end{eqnarray}
Note that $\epsilon_-$ is different from $\epsilon_+$. Although $u_\mathrm{m}^2/u^2-1 = ( u_\mathrm{m}/u -1 )( u_\mathrm{m}/u + 1) \approx 2( u_\mathrm{m}/u -1 ) $ can be used to make $\epsilon_-$ and $\epsilon_+$ similar in form, this approximation will introduce greater error, which has been discussed in Ref.~\cite{Shaikh2019PRD99.104040}.
From Eq.~\eqref{Delta_phi_rm_in}, we can obtain the impact parameter of the strongly deflected photon inside the photon sphere as
\begin{equation}\label{u_PS_in}
u_- = u_\mathrm{m}\left[1+\sqrt{z_\mathrm{O}z_\mathrm{S}}e^{\frac{\bar{b}_-+\pi-\Delta\phi}{\bar{a}_-}}\right]^{-\frac{1}{2}}.
\end{equation}

In the following sections, we will use Eqs.~\eqref{u_PS_out} and \eqref{u_PS_in} to study the shape, thickness and interferometric pattern of the higher-order photon rings.

\section{Inner and outer photon rings in the strong deflection limit}
\label{sec:chapter3}

Considering a thin circular emission disk on the equatorial plane of an ultracompact object,  the strongly deflected photons will form photon rings inside and outside the photon sphere as they are received by the observer. 
The inner photon rings are generally absent in a black hole spacetime and thus unique for an ultracompact object, as numerically shown in Refs.~\cite{Olmo2021PLB829.137045,Gyulchev2021EPJC81.885,Rosa2023PRD107.084048}.
The order of a photon ring can be denoted by the number $n$ of half orbits \cite{Gralla2019PRD100.024018}. 
For $n\geqslant2$, the photon wind around the ultracompact object at least once, passing through the equatorial plane more than twice, generating the so-called higher-order photon rings \cite{Tsupko2022PRD106.064033}.

\subsection{Shape}

The change in the azimuthal angle of the higher-order photon rings was found to be \cite{Tsupko2022PRD106.064033}
\begin{eqnarray} \label{Delta_phi_range}
\Delta\phi &=& \bigg\{ 
        \begin{aligned}
        &\ n\pi+\gamma, & \mathrm{for\ n} = 2,4,6,\cdots \\
        &\ (n+1)\pi-\gamma, &\ \ \ \ \mathrm{for\ n} = 3,5,7,\cdots
        \end{aligned}
\end{eqnarray}
where $\gamma$ is the angle between the observer and the emitter in the ultracompact object-centered reference frame, given by \cite{Luminet1979AA75.228}
\begin{eqnarray}
\cos\gamma &=& \left\{ 
        \begin{aligned}
        &\ \frac{\cos\alpha}{\sqrt{\cos^2\alpha+\cot^2\vartheta_\mathrm{O}}}, & \mathrm{for}\ n = 2,4,6,\cdots \\
        &\ \frac{-\cos\alpha}{\sqrt{\cos^2\alpha+\cot^2\vartheta_\mathrm{O}}}, &\ \ \ \ \mathrm{for}\ n = 3,5,7,\cdots
        \end{aligned}\right.
\end{eqnarray}
with $\vartheta_\mathrm{O}$ being the inclination angle defined according to Ref.~\cite{Tsupko2022PRD106.064033}. 
Here $\alpha$ is not the polar angle $\varphi$ that is normally defined counterclockwise with respect to the horizontal axis \cite{Luminet1979AA75.228}, these two angles satisfy the following relation \cite{Tsupko2022PRD106.064033}
\begin{equation}
    \cos\alpha=-\sin\varphi.
\end{equation}
Substituting the above expressions into Eqs.~\eqref{u_PS_out} and \eqref{u_PS_in}, we can obtain the impact parameter of the $n$th-order outer photon ring as
\begin{eqnarray}
    \label{u_out: for high order photon ring}
    u_{+n}(\varphi) &=& u_\mathrm{m}\bigg[1+f_+(r_\mathrm{S})\mathcal{E}_{+n}(\varphi, \vartheta_\mathrm{O})\bigg],
\end{eqnarray}
and the impact parameter of the $n$th-order inner photon rings as
\begin{eqnarray}
    \label{u_in: for high order photon ring}
    u_{-n}(\varphi) &=& u_\mathrm{m}\bigg[1+ f_-(r_\mathrm{S})\mathcal{E}_{-n}(\varphi, \vartheta_\mathrm{O})\bigg]^{-\frac{1}{2}},
\end{eqnarray}
where
\begin{eqnarray}
    f_+(r_\mathrm{S}) &=& \left(1-\frac{r_\mathrm{m}}{r_\mathrm{O}}\right)\left(1-\frac{r_\mathrm{m}}{r_\mathrm{S}}\right)\mathrm{exp}\left[\frac{\bar{b}_+(r_\mathrm{S})}{\bar{a}_+}\right],\\
    f_-(r_\mathrm{S}) &=& \sqrt{\left(1-\frac{r_\mathrm{m}}{r_\mathrm{O}}\right)\left(1-\frac{r_\mathrm{m}}{r_\mathrm{S}}\right)}\mathrm{exp}\left[\frac{\bar{b}_-(r_\mathrm{S})}{\bar{a}_-}\right], 
\end{eqnarray}
and
\begin{eqnarray}
    \mathcal{E}_{\pm n}(\varphi, \vartheta_\mathrm{O}) &=& \mathrm{exp}\bigg( \frac{1}{\bar{a}_\pm}\arccos\frac{\sin\varphi}{\sqrt{\sin^2\varphi+\cot^2\vartheta_\mathrm{O}}} \bigg)  \nonumber\\
    && \times \mathrm{exp}\bigg(-\frac{n\pi}{\bar{a}_\pm}\bigg) . 
\end{eqnarray}
 Eq.~\eqref{u_out: for high order photon ring} agrees with Eq.~(26) in Ref.~\cite{Aratore2024PRD109.124057}, as both employ the same methods.
Note that $f(r_\mathrm{S})$ used in Ref.~\cite{Tsupko2022PRD106.064033} is only valid for the Schwarzschild black hole, while Eq.~\eqref{u_out: for high order photon ring} is applicable to arbitrary black hole. 
By varying $\varphi$ in the range $[0,2\pi]$, we can obtain the shape of the inner and outer photon rings from Eqs.~\eqref{u_in: for high order photon ring} and \eqref{u_out: for high order photon ring}, respectively.

\subsection{Thickness}

When the emission disk have an inner boundary $r_\mathrm{S}^\mathrm{in}$ and outer boundary $r_\mathrm{S}^\mathrm{out}$, the resulting inner or outer photon rings will possess a certain thickness $\Delta u_n$.
By adopting Eqs.~\eqref{u_out: for high order photon ring} and \eqref{u_in: for high order photon ring}, we can obtain
\begin{eqnarray}
    \label{Delta_u_out: for high order photon ring}
    \Delta u_{+n}(\varphi) &\equiv& \bigg|u^\mathrm{out}_{+n}(\varphi)-u^\mathrm{in}_{+n}(\varphi)\bigg| \nonumber \\
    &=& u_\mathrm{m}\left[f_+(r_\mathrm{S}^\mathrm{out})-f_+(r_\mathrm{S}^\mathrm{in})\right]\mathcal{E}_{+n}(\varphi, \vartheta_\mathrm{O})
\end{eqnarray}
for the $n$th-order outer rings, and 
\begin{eqnarray}
    \label{Delta_u_in: for high order photon ring}
    \Delta u_{-n}(\varphi) &\equiv& \bigg|u^\mathrm{out}_{-n}(\varphi)-u^\mathrm{in}_{-n}(\varphi)\bigg| \nonumber \\
    &=& u_\mathrm{m}\bigg\{\bigg[1+f_-(r_\mathrm{S}^\mathrm{in})\mathcal{E}_{-n}(\varphi, \vartheta_\mathrm{O})\bigg]^{-\frac{1}{2}}\nonumber\\
    &&-\bigg[1+f_-(r_\mathrm{S}^\mathrm{out})\mathcal{E}_{-n}(\varphi, \vartheta_\mathrm{O})\bigg]^{-\frac{1}{2}}\bigg\}
\end{eqnarray}
for the $n$th-order inner rings.  Eq.~\eqref{Delta_u_out: for high order photon ring} is a direct generalization of corresponding equations in Ref.~\cite{Tsupko2022PRD106.064033}.

\section{Interferometric pattern of the inner and outer photon rings}
\label{sec:chapter4}

Higher-order photon rings can be detected by very long baseline interferometry \cite{Johnson2020SciAdv6.eaaz1310}. 
The interferometric signature of a photon ring is its visibility, defined as the following Fourier transform \cite{Thompson2017Book}
\begin{equation}\label{def:Visibility}
    \mathcal{V}(\vec{\mathrm{u}})=\int I(\vec{r})e^{-j2\pi\vec{\mathrm{u}}\cdot\vec{r}}\mathrm{d}^2\vec{r},
\end{equation}
where $I(\vec{r})$ is the intensity distribution of the photon ring, $\vec{r} = (x, y)$ is the sky position in radians, $\vec{\mathrm{u}}$ is the baseline vector in units of the observed wavelength $\lambda$. 

The calculation of $\mathcal{V}(\vec{\mathrm{u}})$ in principle requires a complete two-dimensional Fourier transform that involves huge computational costs. 
However, from the perspective of image reconstruction, instead we can first cut one-dimensional slices of the visibility in different directions and then combine them to get a two-dimensional visibility, which is just the basic idea of the projection-slice theorem \cite{Thompson2017Book}. 

In order to apply this theorem, we denote $\vec{\mathrm{u}}=(\mathrm{u}, \sigma)$ in polar coordinates, where $\sigma$ is a projection angle and $\mathrm{u}$ is the projected length of the baseline in the direction of $\sigma$, and rotate the projection axis clockwise by the angle of $\sigma$ to make it horizontal for mathematical convenience. Then the sky coordinates of a photon ring can be written as 
\begin{eqnarray}\label{eq: x and y}
    x(\varphi) &=& u(\varphi)\cos(\varphi-\sigma),\\
    y(\varphi) &=& u(\varphi)\sin(\varphi-\sigma).
\end{eqnarray}
Here we omit the subscript $``\pm"$ of $u(\varphi)$ and  notice that the following calculations are applied for both the inner and outer photon rings. We also note that in order to obtain $\vec{r} = (x, y)$ in radians, dividing $u$ by the distance from the lens to the observer is required.

In general, we can use the arclength $\tilde{s}$ to parametrize the shape curve of the photon ring as $u(\varphi)=u(\tilde{s})=( x(\tilde{s}), y(\tilde{s}) )$. 
When the baseline is not long enough to resolve the thickness of a photon ring, i.e.,
\begin{eqnarray}\label{regime_considered}
    \mathrm{u} \ll (\Delta u)^{-1},
\end{eqnarray}
the projection of the intensity distribution on the horizontal axis may be given by \cite{Gralla2020PRD102.044017}
\begin{eqnarray}\label{eq: PxI}
    \mathcal{P}_x I(x) &=& \int \mathcal{I}(\tilde{s})\delta(x-x_0(\tilde{s}))\mathrm{d}\tilde{s}\nonumber\\
    &=& \sum_{x=x_0} \frac{\mathcal{I}}{\mathrm{d}x_0/\mathrm{d}\tilde{s}}.
\end{eqnarray}
where $\mathcal{I}(\tilde{s})$ is the integrated intensity, for a photon ring with uniform brightness, we have $\mathcal{I}(\tilde{s})=I_0=I_\mathrm{tot}/\int\mathrm{d}\tilde{s}$.
$x_0$ is the point where the vertical line intersects the shape curve.

For $\mathrm{d}x_0/\mathrm{d}\tilde{s}=0$, the vertical line is tangent to the shape curve, we denote the corresponding $x_0$ as the tangential point $x_\mathrm{T}$ (i.e., the vertical point of tangency). From Eq.~\eqref{eq: PxI},  we know that the projection intensity $\mathcal{P}_x I(x)$ becomes singular at $x = x_\mathrm{T}$, which implies that the behavior of the resulting visibility is dominated by the tangential points \cite{Gralla2020PRD102.044017}.

In order to find out the positions of all the tangential points on the shape curve, we rewrite the infinitesimal arclength as
\begin{eqnarray}
    \mathrm{d}\tilde{s} &=& \sqrt{x'(\varphi)^2+y'(\varphi)^2}\ \mathrm{d}\varphi \nonumber\\
    &=&\sqrt{u(\varphi)^2+u'(\varphi)^2}\ \mathrm{d}\varphi\ne0,
\end{eqnarray}
where the prime stands for the derivative with respect to $\varphi$. 
Meanwhile, by adopting the chain rule, we have
\begin{eqnarray}
    \frac{\mathrm{d}x_0}{\mathrm{d}s}=\frac{\mathrm{d}x_0}{\mathrm{d}\varphi}\frac{\mathrm{d}\varphi}{\mathrm{d}s}.
\end{eqnarray}
Then the condition of $\mathrm{d}x_0/\mathrm{d}\tilde{s}=0$ can be converted to $\frac{\mathrm{d}x_0}{\mathrm{d}\varphi}\big|_{x_0=x_\mathrm{T}} = 0$, and with Eq.~\eqref{eq: x and y} we can have
\begin{eqnarray}\label{eq: varphi_T}
    \mathcal{F}(\varphi_\mathrm{T}) &=& u'(\varphi_\mathrm{T})\cos(\varphi_\mathrm{T}-\sigma) - u(\varphi_\mathrm{T})\sin(\varphi_\mathrm{T}-\sigma) \nonumber\\
    &=& 0,
\end{eqnarray}
where $\varphi_\mathrm{T}$ is the corresponding polar angle of $x_\mathrm{T}$ and its value can be obtained from Eq.~\eqref{eq: varphi_T} by employing a FindRoot algorithm. 

From Eq.~\eqref{eq: x and y} we can further obtain $x_\mathrm{T}$.
Then $\mathcal{P}_x I(x)$ in the small two-sided neighborhood of $x_\mathrm{T}$ can be approximated by the Heaviside step functions (see Eq.~(28) in Ref.~\cite{Gralla2020PRD102.044017}). 
The Fourier transfrom of the approximated $\mathcal{P}_x I(x)$ gives the following visibility \cite{Gralla2020PRD102.044017}
\begin{eqnarray}\label{Visibility: GAV}
    \mathcal{V}(\mathrm{u},\sigma) &=& \sum_{x=x_\mathrm{T}}\mathcal{I}_\mathrm{T}\sqrt{\mathcal{R}_\mathrm{T}}\ e^{-j\frac{\pi}{4}S_\mathrm{T}}\frac{e^{-j2\pi\mathcal{Z}_\mathrm{T}\mathrm{u}}}{\sqrt{\mathrm{u}}},
\end{eqnarray}
where the subscript $``\mathrm{T}"$ denotes quantities evaluated at $x=x_\mathrm{T}$.
$\mathcal{I}_\mathrm{T}$ is the integrated intensity.
$S_\mathrm{T}=\pm1$ depending on whether the horizontal axis and the inward normal vector (pointing to the curvature center) at the tangential point are in the same $(+1)$ or opposite $(-1)$ direction.
$\mathcal{Z}_\mathrm{T}$ is the projected position of the tangential point on the horizontal axis.
According to the definition in Ref.~\cite{Gralla2020PRD102.044017}, the projected position of a point on the shape curve can be written as
\begin{eqnarray}\label{def:projection position}
   \mathcal{Z}(\sigma') = x(\varphi)\cos(\sigma')+y(\varphi)\sin(\sigma').
\end{eqnarray}
For the horizontal projection axis, we have $\sigma'=0$ and thus obtain 
\begin{eqnarray}\label{z_T}
    \mathcal{Z}_\mathrm{T}=x_\mathrm{T}(\varphi_\mathrm{T}).
\end{eqnarray}
$\mathcal{R}_\mathrm{T}$ is the curvature radius \cite{Gralla2020PRD102.044017}, given by 
\begin{eqnarray}\label{def:curvature radius}
    \mathcal{R}_\mathrm{T} = \bigg|\sqrt{[x'(\varphi)]^2+[y'(\varphi)]^2}\ \frac{\mathrm{d}\varphi}{\mathrm{d}\sigma'}\bigg|_{\varphi=\varphi_\mathrm{T}}.
\end{eqnarray}
By taking the derivative of Eq.~\eqref{def:projection position} with respect to $\varphi$, we can obtain \cite{Gralla2020PRD102.124003}
\begin{eqnarray}\label{relation: varphi_nor vs varphi}
    \tan(\sigma')=-\frac{x'(\varphi)}{y'(\varphi)},
\end{eqnarray}
which allows us to further compute 
\begin{eqnarray}\label{R_T}
    \mathcal{R}_\mathrm{T} = \frac{ \left[y'(\varphi_\mathrm{T}) \right]^2}{\big| x''(\varphi_\mathrm{T}) \big|}.
\end{eqnarray}
Here we have used the condition of $x'(\varphi_\mathrm{T})=0$. 

To sum up, in the regime \eqref{regime_considered}, the final expression \eqref{Visibility: GAV} for the visibility of the outer or inner higher-order photon rings is proposed by Ref.~\cite{Gralla2020PRD102.044017}. By adopting the strong deflection limit method, we find that the coefficients included can be simply evaluated with the help of Eqs.~\eqref{eq: varphi_T}\eqref{z_T} and \eqref{R_T}.

\section{Applications to specific ultracompact object}
\label{sec:chapter5}

In this work, we take the Schwarzschild star with a photon sphere as a simple example, while our formulae can be applied to other ultracompact objects. 

The Schwarzschild star is an isotropic self-gravitating object with a uniform energy density, being an exact solution of general relativity \cite{Stephani2009book}.
Although sufferring from the Buchdahl limit \cite{Buchdahl1959PR116.1027} and the oversimplified uniform matter distribution, it has still attracted great attention, such as its anisotropy \cite{Gabbanelli2019EPJC79.486, Ovalle2019CQG36.205010}, time dependence \cite{Beltracchi2019PRD99.084021}, connection to the gravastar \cite{Mazur2023Universe9.88,Mazur2004PNAS101.9545}, nonvanishing positive tidal Love number \cite{Chirenti2020CQG37.195017} and power-law tail very similar to the Schwarzschild black hole in the gravitational wave ringdown waveform \cite{Konoplya2019PRD100.044027}. 
Besides, the Schwarzschild star may be the simplest model to study the common features of ultracompact objects.

The explicit form of the Schwarzschild star metric is given by ($G=c=1$) \cite{Schwarzschild1916SPAWB424,Wald1984Book,Shaikh2019PRD99.104040,Gao2024PRD109.063030}
\begin{eqnarray}\label{metric: SchStar}
A(r) &=& \left\{ 
        \begin{aligned} & \left(\frac{3}{2}\sqrt{ \mathcal{H}_R}-\frac{1}{2}\sqrt{\mathcal{H}_r}\right)^2, & r<R\\
        &\ 1-\frac{R_\mathrm{s}}{r}, & r\geqslant R
        \end{aligned}
        \right., \\
B(r) &=& \left\{ 
        \begin{aligned}
        &\ \mathcal{H}_r^{-1}, & \phantom{\int XXX} r<R\\
        &\left(1-\frac{R_\mathrm{s}}{r}\right)^{-1}, & r\geqslant R
        \end{aligned}
        \right.,\\
C(r) &=& r^2,
\end{eqnarray}
where $R_\mathrm{s}=2M$ is the Schwarzschild radius, $M$ is the ADM mass. $R$ is the radius of the Schwarzschild star, its value is constrained by the Buchdahl's theorem \cite{Buchdahl1959PR116.1027}. 
$\mathcal{H}_R$ and $\mathcal{H}_r$ are respectively defined as
\begin{eqnarray} 
      \label{def: H_R}
      \mathcal{H}_R &=& 1-\frac{R_\mathrm{s}}{R},\\
      \label{def: H_r}
      \mathcal{H}_r &=& 1-\frac{R_\mathrm{s}}{R^3}r^2.
    \end{eqnarray} 

We assume that the \SchStar does not emit electromagnetic waves itself and has an electromagnetically transparent surface with relatively high absorption rate in its interior \cite{Cardoso2019LRR22.4,Gao2024PRD109.063030,Sakai2014PRD90.104013}, then its optical appearance may have a dark spot in the center that is consistent with current observed images taken by the  Event Horizon Telescope \cite{EHTC2019ApJ875.L6,EHTC2022ApJ930.L17}.
More importantly, in this scenario photons can enter the region where $r < R$, providing the possibility for the formation of the inner photon rings.

For a Schwarzschild star with $9R_\mathrm{s}/8 < R < 3R_\mathrm{s}/2$, it is shown that the geodesic equation \eqref{GeodesicEquationOfSchStar} is well-defined under the above assumptions, and the lensed photons entering the photon sphere could reach an observer \cite{Gao2024PRD109.063030}. 
In this configuration,
the photon sphere locates at
\begin{equation}
    r_\mathrm{m}=\frac{3}{2}R_\mathrm{s},
\end{equation}
and the antiphoton sphere locates at  \cite{Gao2024PRD109.063030}
\begin{equation}\label{APSOfSchStar}
    r_\mathrm{a}=\frac{(R/R_\mathrm{s})^{3/2}}{3}\sqrt{\frac{8R/R_\mathrm{s}-9}{R/R_\mathrm{s}-1}}R_\mathrm{s}.
\end{equation}
From Eq.~\eqref{rcdef}, we can also find \cite{Gao2024PRD109.063030}
\begin{eqnarray}\label{rcOfSchStar}
    r_\mathrm{c} &=& \frac{3\sqrt{3}R}{16(R/R_\mathrm{s})^3+27}\bigg[12\sqrt{(R/R_\mathrm{s})^4-(R/R_\mathrm{s})^3} \nonumber\\
    &&-\sqrt{16(R/R_\mathrm{s})^4-216(R/R_\mathrm{s})+243}\bigg].
\end{eqnarray}

By taking the \SchStar with $R=2.5M$ as an example, we first evaluate the accuracy of Eqs.~\eqref{u_PS_out} and ~\eqref{u_PS_in}. Their error with respect to numerical results increases as $\Delta\phi$ decrease, and the relative error of $u_-$ is much larger than that of $u_+$. For $\Delta\phi\geqslant2\pi$ (that is always satisfied for higher-order photon rings, see Eq.~\eqref{Delta_phi_range}), we find that $u_-$ has the largest relative error that is about 7\% when $\Delta\phi=2\pi$, but this value reduces to $0.6\%$ when $\Delta\phi$ increases to $3\pi$. These errors are reasonably accepted in this work. 

\begin{figure}[H]    
	\centering
	\includegraphics[width=1\textwidth]{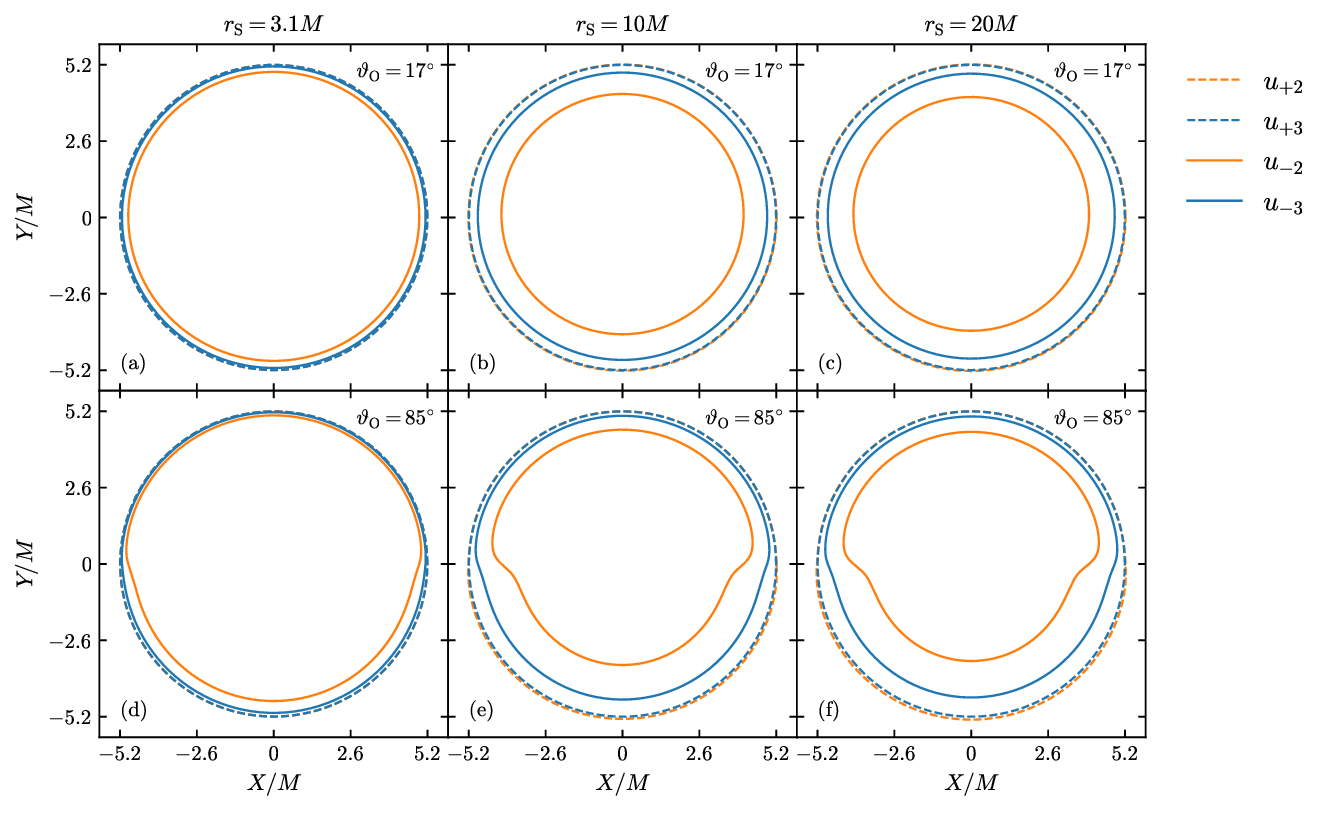}
	\caption{The shape curves of the $n=2, 3$ inner and outer photon rings of the Schwarzschild star with radius $R=2.5M$. From the first column to the third column, the radius $r_\mathrm{S}$ of the thin circular emission disk is $3.1M, 10M, 20M$, respectively. The inclination angle $\vartheta_\mathrm{O}$ is $17^\circ$ in the first row, and $85^\circ$ in the second row.}
	\label{Fig1:PhotonRingShape}
\end{figure}

With Eqs.~\eqref{u_in: for high order photon ring} and \eqref{u_out: for high order photon ring}, we show the shape curves of its $n=2, 3$ inner and outer photon rings in Fig.~\ref{Fig1:PhotonRingShape}.
It can be seen that the $n=2$ inner ring is nearly circular for a low inclination ($\vartheta_\mathrm{O}=17^\circ$),  but its lower half can get clearly deformed for a high inclination ($\vartheta_\mathrm{O}=85^\circ$). 
The same is true for the  $n=3$ inner ring, while for the $n=2$ or 3 outer ring, the shape curve is always nearly circular. The term ``nearly" is adopted here because when the inclination angle is high, the lower half of the outer photon ring can deviate from a circular shape slightly.
Fig.~\ref{Fig1:PhotonRingShape} also shows that the inner photon rings gradually approach the image center as the radius $r_\mathrm{S}$ of the emission disk increases.

\begin{figure}[htbp]    
	\centering
	\includegraphics[width=0.6\textwidth]{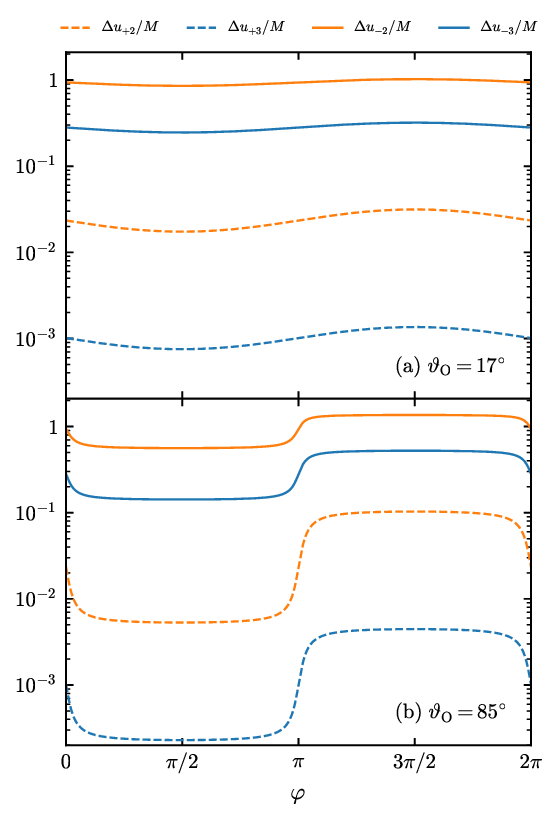}
	\caption{ The thicknesses of the $n=2, 3$ inner and outer photon rings of the Schwarzschild star with radius $R=2.5M$, given a thin circular emission disk with inner boundary $r_\mathrm{S}^\mathrm{in}=3.1M$ and outer boundary $r_\mathrm{S}^\mathrm{out}=20M$. (a) The inclination angle $\vartheta_\mathrm{O}=17^\circ$, (b) $\vartheta_\mathrm{O}=85^\circ$.}
	\label{Fig2:PhotonRingThickness}
\end{figure}

Then we use Eqs.~\eqref{Delta_u_in: for high order photon ring} and \eqref{Delta_u_out: for high order photon ring} to obtain the thicknesses of the $n=2, 3$ inner and outer photon rings of the \SchStar with $R = 2.5M$.
Results are shown in Fig.~\ref{Fig2:PhotonRingThickness}.
Here we set $r_\mathrm{S}^\mathrm{in}=3.1M$ and $r_\mathrm{S}^\mathrm{out}=20M$, and show the thicknesses of the $n=2, 3$ inner and outer photon rings of the \SchStar with $R = 2.5M$.
We find that the thicknesses of these photon rings satisfy the relation of $\Delta u_{-2}>\Delta u_{-3}>\Delta u_{+2}>\Delta u_{+3}$, and the thickness of the inner rings has a maximum at $\varphi=3\pi/2$, similar to that of the outer ones \cite{Tsupko2022PRD106.064033}.

\begin{figure}[htbp]    
	\centering
	\includegraphics[width=0.6\textwidth]{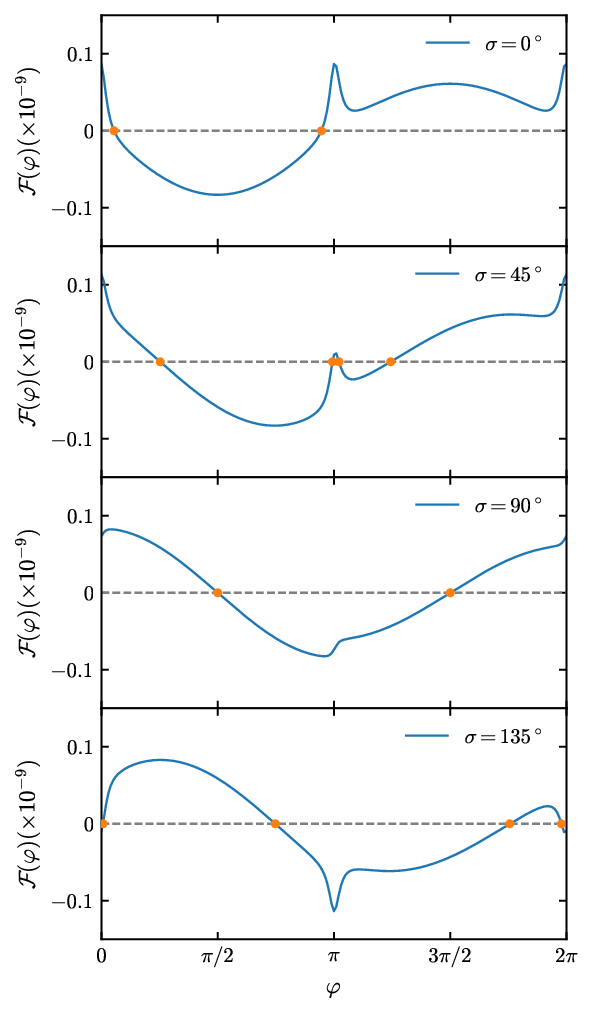}
	\caption{ $\mathcal{F}(\varphi)$ of the $n=2$ inner photon ring. From top to bottom, the projection angle $\sigma$ is respectively $0^\circ, 45^\circ, 90^\circ, 135^\circ$. The gravitational lens is assumed to be a $R=2.5M$ \SchStar with mass and distance equal to M87*. The light source is a thin circular emission disk with radius $r_\mathrm{S} = 20M$ and inclination $\vartheta_\mathrm{O} = 85^\circ$. The orange solid dots mark $\varphi_\mathrm{T}$ of the tangential points.}
	\label{Fig3:TangentialPoints}
\end{figure}

\begin{figure}[t!]    
	\centering
	\includegraphics[width=1.0\textwidth]{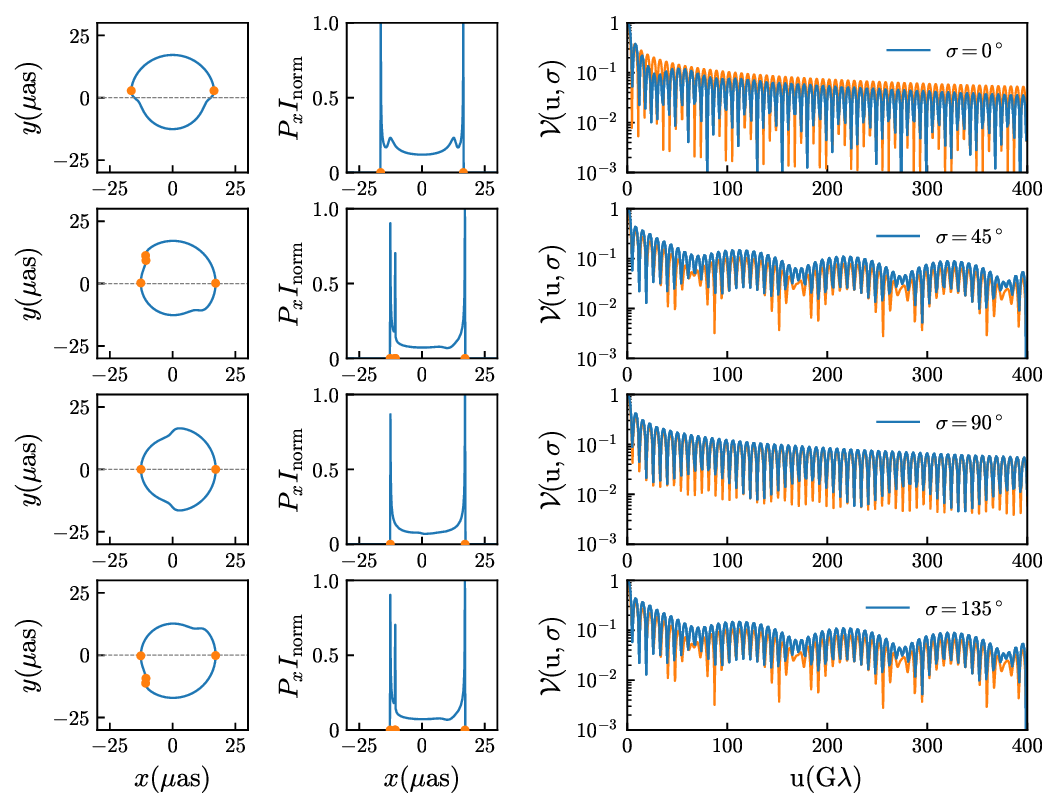}
	\caption{The shape curve, normalized projected intensity $P_x I_\mathrm{norm}$ and (normalized) visibility $\mathcal{V}(\mathrm{u}, \sigma)$ of the $n=2$ inner photon ring. The lens is assumed to be a $R=2.5M$ \SchStar with mass and distance equal to M87*. The source is a thin circular emission disk with radius $r_\mathrm{S} = 20M$ and inclination $\vartheta_\mathrm{O} = 85^\circ$. The blue solid lines represent numerical results (Radon transform for $P_x I_\mathrm{norm}$ and Fourier transform for $\mathcal{V}$). The orange solid dots mark the positions of the tangential points. The orange solid line represents the visibility given by our formulae.}
	\label{Fig4:PhotonRing_Inside_Visibility}
\end{figure}

\begin{figure}[t!]    
	\centering
	\includegraphics[width=1.0\textwidth]{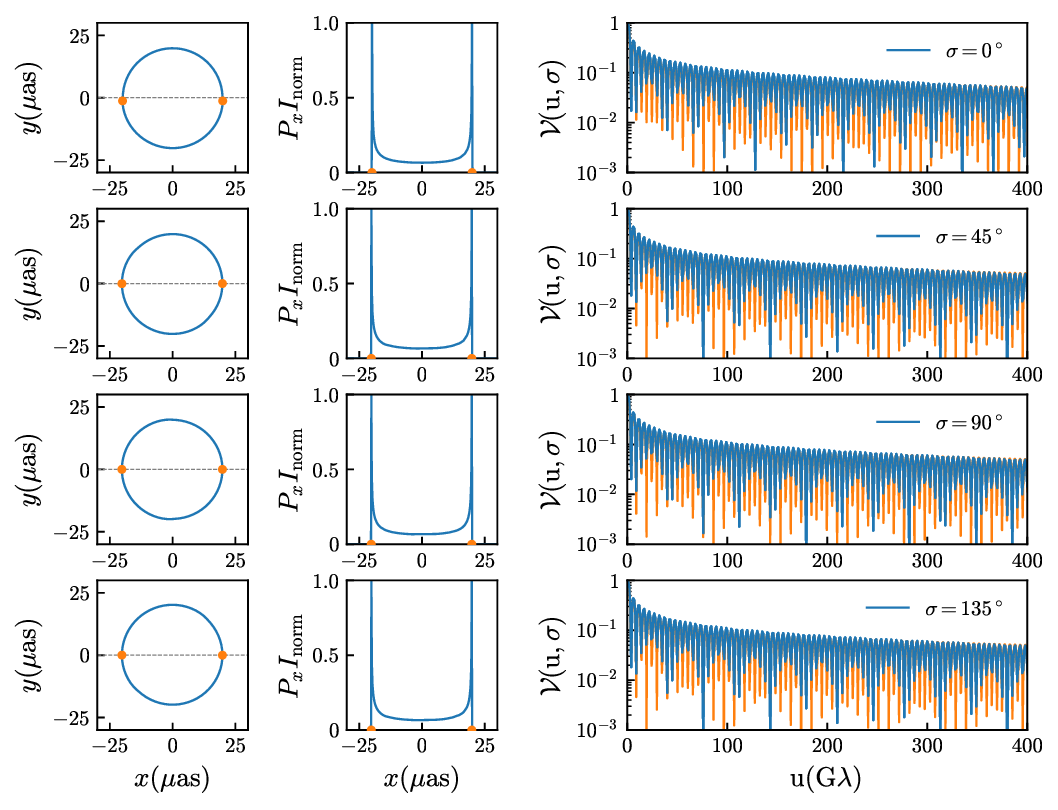}
	\caption{The shape curve, normalized projected intensity $P_x I_\mathrm{norm}$ and (normalized) visibility $\mathcal{V}(\mathrm{u}, \sigma)$ of the $n=2$ outer photon ring. The physical setup and color representation are the same as in Fig.~\ref{Fig4:PhotonRing_Inside_Visibility}. }
	\label{Fig5:PhotonRing_Outside_Visibility}
\end{figure}

To compute the visibility of the higher-order inner and outer photon rings, we first need to determine the location of the tangential points on the shape curve by using Eq.~\eqref{eq: varphi_T}.
Assuming that the mass and distance of the \SchStar with $R=2.5M$ are the same as M87* \cite{EHTC2019ApJ875.L6}, and the inclination of the surrounding thin circular disk with $r_\mathrm{S}=20M$ is $\vartheta_\mathrm{O} = 85^\circ$, we show $\mathcal{F}(\varphi)$ of the $n=2$ inner photon ring in Fig.~\ref{Fig3:TangentialPoints}. 
The roots of  $\mathcal{F}(\varphi)$ give $\varphi_\mathrm{T}$ of the tangential points, which are marked by the orange solid dots.
We find that the number of the tangential points is 2 for $\sigma = 0^\circ$ or $90^\circ$, while the number is 4 for $\sigma = 45^\circ$ or $135^\circ$. 
The locations of these tangential points on the shape curves are shown in the first column of Fig.~\ref{Fig4:PhotonRing_Inside_Visibility}.
We also find that the number of the tangential points of the $n=2$ outer photon ring is always 2 and show their locations on the shape curves in the first column of Fig.~\ref{Fig5:PhotonRing_Outside_Visibility}.

In the second column of Fig.~\ref{Fig4:PhotonRing_Inside_Visibility} and~\ref{Fig5:PhotonRing_Outside_Visibility}, for each $\sigma$ we use the blue solid line to show the normalized projected intensity $P_x I_\mathrm{norm} = P_x I / (P_x I)_\mathrm{max}$, which is obtained by adopting a numerical Radon transform algorithm \cite{Paugnat2022AAp668.A11}, i.e., the complete integral of the intensity $I$ along the vertical lines. On the horizontal axis, we use the orange solid dots to mark the positions of $x_\mathrm{T}$. It can be seen that $P_x I_\mathrm{norm}$ indeed has a singular behavior at $x = x_\mathrm{T}$, which agrees with the previous discussion. 
Besides, we find that $\mathcal{P}_x I_\mathrm{norm}$ of the $n=2$ inner photon ring has additional substructures in $|x|<|x_\mathrm{T}|$ for $\sigma = 0^\circ$, which may produce a subdominant term of the visibility and are beyond the prediction of  Eq.~\eqref{Visibility: GAV}.

In the third column of Fig.~\ref{Fig4:PhotonRing_Inside_Visibility} and~\ref{Fig5:PhotonRing_Outside_Visibility}, we use the blue solid line to show the (normalized) visibility $\mathcal{V}(\mathrm{u}, \sigma)$ corresponding to the numerical $P_x I_\mathrm{norm}$ for each $\sigma$, and use the orange solid line to show the (normalized) visibility given by our formulae. 
It is shown that the oscillating phase of $\mathcal{V}(\mathrm{u},\sigma)$ given by our formulae coincides exactly with that of the numerical $\mathcal{V}(\mathrm{u}, \sigma)$, while the amplitude of it can be different from that of the numerical one, as illustrated in the first, second and fourth rows of Fig.~\ref{Fig4:PhotonRing_Inside_Visibility}. The reason may include the following two aspects.
First, the $n=2$ inner photon ring can have substructures that are not contained in Eq.~\eqref{Visibility: GAV} (for the case of $\sigma = 0^\circ$), producing unpredicted additional interference envelopes in $\mathcal{V}(\mathrm{u},\sigma)$.
Second, the numerical algorithm undersamples in the neighborhood of $x_\mathrm{T}$ (for the case of $\sigma = 45^\circ$ and $135^\circ$), making the integral intensity differ from the prediction of Eq.~\eqref{Visibility: GAV} and further affecting the normalization of $\mathcal{V}(\mathrm{u},\sigma)$.
As we shall see below, these discrepancies get reduced when the overall visibility of multiple photon rings is considered, which indicates that the superposition of singular tangential points from different photon rings can weaken the effects of substructures and undersampling.

Based on the current observational results of M87* and Sgr A*, their photon rings should be nearly circular since the observational inclinations are both estimated to be low \cite{EHTC2022ApJ930.L16}. 
For these two objects, the number of the tangential points would always be 2 and there would be no distinct additional substructures in the intensity profile.  
Therefore, we expect that $\mathcal{V}(\mathrm{u},\sigma)$ given by Eq.~\eqref{Visibility: GAV} can approximate the actual visibility of the higher-order photon rings very well. 

From Fig.~\ref{Fig4:PhotonRing_Inside_Visibility}, we also find that $\mathcal{V}(\mathrm{u}, \sigma)$ of the $n=2$ inner photon ring may have different characteristics depending on $\sigma$. 
For $\sigma = 0^\circ$ and $90^\circ$, the only two tangential points lead to the fact that $\mathcal{P}_x I_\mathrm{norm}$ is very similar to that of a circular ring, thus the resulting $\mathcal{V}(\mathrm{u},\sigma)$ behaves like a cosine function.
For $\sigma = 45^\circ$ and $135^\circ$, 
distinct envelopes may appear due to the presence of 4 tangential points, which are absent in $\mathcal{V}(\mathrm{u},\sigma)$ of the outer photon ring and provide some evidence for the existence of the Schwarzchild star.
As we can see, the first envelope appears at a position of approximately $100\mathrm{G}\lambda$.
Given the fact that the intensity of the photon rings decreases exponentially as the order 
$ n $ increases, and the width effect causes the visibility modulus to exhibit a staircase-like shape where lower-order photon rings dominate higher steps \cite{Johnson2020SciAdv6.eaaz1310}, the observability of the $n=2$ inner photon ring  at $100\mathrm{G}\lambda$ will be crucial for distinguishing the ultracompact object from the black hole.
Although the current observational capability of the Event Horizon Telescope is around $10\mathrm{G}\lambda$ \cite{EHTC2022ApJ930.L12}, in the future it is highly promising to achieve resolutions corresponding to several tens to 100 $\mathrm{G}\lambda$, with the frequency upgrading of the next-generation Event Horizon Telescope \cite{Vincent2022AAp667.A170, Paugnat2022AAp668.A11,Roelofs2022Gal11.12}, or the realization of the space-borne very long baseline interferometry with a baseline equivalent to the Earth-Moon distance \cite{Johnson2020SciAdv6.eaaz1310,Gao2024PRD109.063030}.
At that time, it will be very likely to distinguish the photon-ring differences between the ultracompact object and the black hole at 100 $\mathrm{G}\lambda$ in the visibility regime.

\subsection{Overall visibility of the inner and outer photon ring pairs}

\begin{figure}[t!]    
	\centering
	\includegraphics[width=1.0\textwidth]{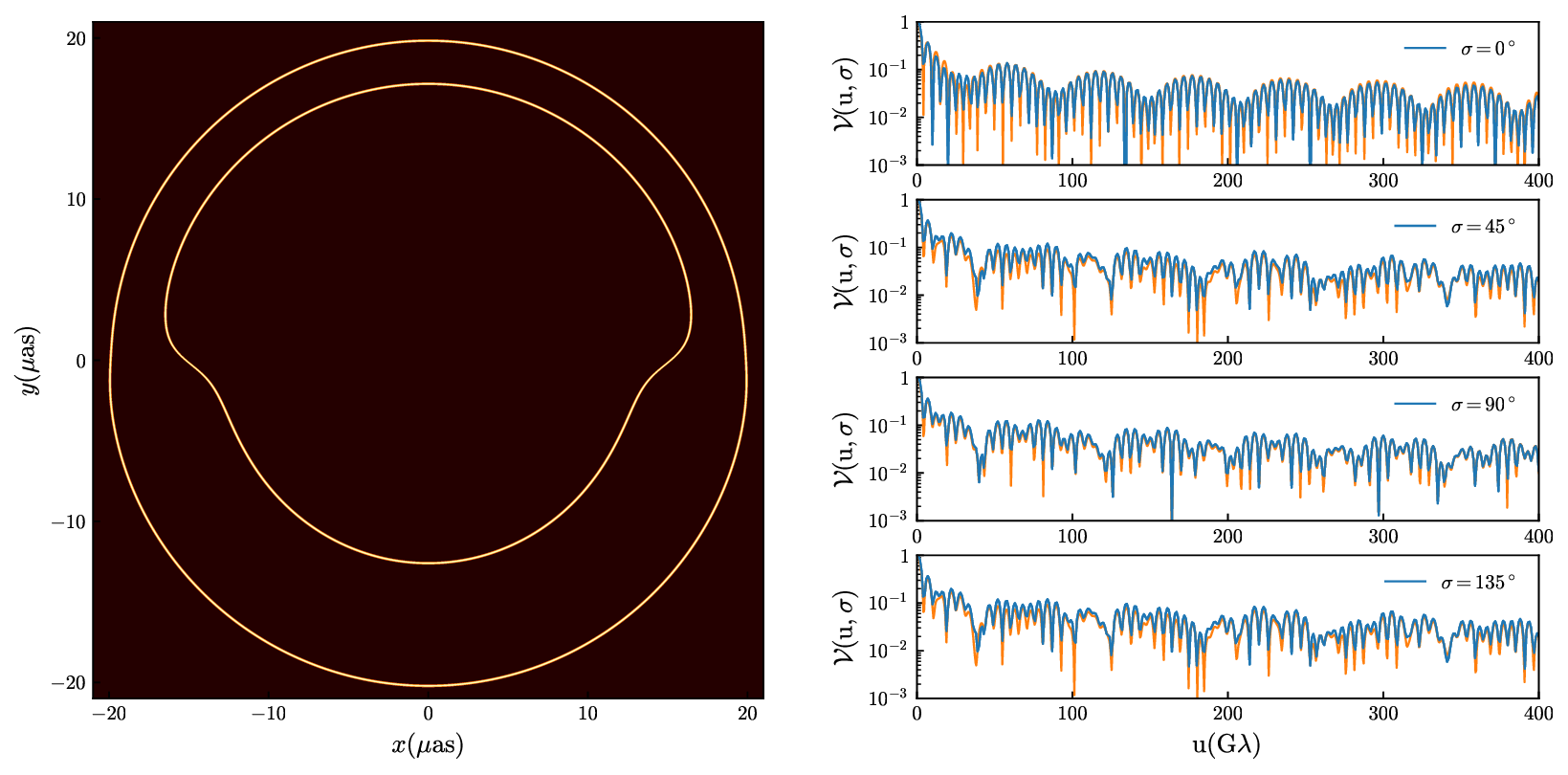}
	\caption{The overall (normalized) visibility $\mathcal{V}(\mathrm{u}, \sigma)$ of the $n=2$ inner and outer photon rings. The physical setup and color representation is the same as in Fig.~\ref{Fig4:PhotonRing_Inside_Visibility}. The integrated intensities of the inner and outer rings are assumed to be identical. }
	\label{Fig6:PhotonRing_Together_Visibility}
\end{figure}

An ultracompact object is shown to have at least an unstable photon sphere and a stable antiphoton sphere \cite{Cunha2017PRL119.251102, Guo2020PRD103.104031}, which leads to the fact that Eq.~\eqref{rcdef} can always be satisfied. 
From the point of view of the effective potential, if there are strongly deflected photons outside the photon sphere, there will also be strongly deflected photons inside the photon sphere.
Thus, the inner and outer photon rings will appear simultaneously for an ultracompact object. Motivated by this, we show the overall visibility of the $n=2$ inner and outer photon rings in Fig.~\ref{Fig6:PhotonRing_Together_Visibility}. Here we assume the integrated intensity on the shape curves is equal everywhere, i.e.,  $\mathcal{I}(s)=1$. 
As we can see, the overall visibility given by our formulae are still in good agreement with the numerical ones. 
This good match may imply that by superimposing the tangential points of different photon rings, we can potentially generalize our results to approximate the visibility of a photon ring with thickness.

In addition, we also find that the overall visibility gets more complicated than the individual inner one shown in Fig.~\ref{Fig4:PhotonRing_Inside_Visibility} (or outer one shown in Fig.~\ref{Fig5:PhotonRing_Outside_Visibility}).
However, given that the intensity of an inner ring should actually be quite different from that of an outer ring due to the effect of gravitational lensing and gravitational redshift \cite{Gralla2019PRD100.024018, Shaikh2019PRD99.104040, Zhu2020EPJC80.444, Gao2022EPJC82.162}, and that the thickness of a photon ring may cause its intensity profile to decrease exponentially \cite{Johnson2020SciAdv6.eaaz1310,CardenasAvendano2023PRD108.064043}, the actual overall visibility may get simplified with a staircase-like structure.

Finally, we give $ \{ \mathcal{I}_\mathrm{T}, \mathcal{R}_\mathrm{T}, \mathcal{S}_\mathrm{T}, \mathcal{Z}_\mathrm{T} \} $ for the visibility of the $n=2$ and $3$ photon rings of the Schwarzschild star with radius $R=2.5M$ in Table.~\ref{tb: parameter for photon ring 2} and ~\ref{tb: parameter for photon ring 3}, respectively. In addition, the visibility of the $n=3$ inner and outer photon rings are shown in Appendix.~\ref{App：n3 photon ring}.

\begin{table}[t!]
	\centering   
	\caption{$ \{ \mathcal{I}_\mathrm{T}, \mathcal{R}_\mathrm{T}, \mathcal{S}_\mathrm{T}, \mathcal{Z}_\mathrm{T} \} $ for the visibility of the $n=2$ photon rings of the Schwarzschild star with radius $R=2.5M$.}
	\resizebox{0.8\textwidth}{!}{
	\begin{tabular}{llllcl}
		\hline
		& &\hspace{2.5cm} $\phantom{\int XX} n=2$ &  \\
		& & Inner ring & Outer ring \\
		\hline
		\multirow{4}{*}{$\sigma=0^\circ \phantom{\int XXX}$} 
		& $\mathcal{I}_\mathrm{T} \phantom{\int XXX}$ & $(1, 1)$ & $(1, 1)$ \\
		& $\mathcal{R}_\mathrm{T} (\times10^{-9}) \phantom{\int XXX}$ & $(0.0304, 0.0304)$ & $(0.0791, 0.0791)$ \\
		& $\mathcal{S}_\mathrm{T} \phantom{\int XXX}$ & $(-1, 1)$ & $(1, 1)$ \\
		& $\mathcal{Z}_\mathrm{T} (\times10^{-9}) \phantom{\int XXX}$ & $(0.0798, -0.0798)$ & $(0.0967, -0.0967)$ \\
		\hline
		\multirow{4}{*}{$\sigma=45^\circ$} 
		& $\mathcal{I}_\mathrm{T}$ & $(1, 1, 1, 1)$ & $(1, 1)$ \\
		& $\mathcal{R}_\mathrm{T} (\times10^{-9})$ & $(0.0812, 0.0249, 0.0209, 0.0652)$ & $(0.0962, 0.0973)$ \\
		& $\mathcal{S}_\mathrm{T}$ & $(-1, 1, -1, 1)$ & $(-1, 1)$ \\
		& $\mathcal{Z}_\mathrm{T} (\times10^{-9})$ & $(0.0830, -0.0515, -0.0522, -0.0614) \phantom{\int XXX}$ & $(0.0962, -0.0979)$ \\
		\hline
		\multirow{4}{*}{$\sigma=90^\circ$} 
		& $\mathcal{I}_\mathrm{T}$ & $(1, 1)$ & $(1, 1)$ \\
		& $\mathcal{R}_\mathrm{T} (\times10^{-9})$ & $(0.0827, 0.0619)$ & $(0.0962, 0.0978)$ \\
		& $\mathcal{S}_\mathrm{T}$ & $(-1, 1)$ & $(1, 1)$ \\
		& $\mathcal{Z}_\mathrm{T} (\times10^{-9})$ & $(0.0832, -0.0611)$ & $(0.0962, -0.0980)$ \\
		\hline
		\multirow{4}{*}{$\sigma=135^\circ$} 
		& $\mathcal{I}_\mathrm{T}$ & $(1, 1, 1, 1)$ & $(1, 1)$ \\
		& $\mathcal{R}_\mathrm{T} (\times10^{-9})$ & $(0.0812, 0.0249, 0.0209, 0.0652)$ & $(0.0962, 0.0973)$ \\
		& $\mathcal{S}_\mathrm{T}$ & $(-1,1,-1,1)$ & $(-1, 1)$ \\
		& $\mathcal{Z}_\mathrm{T} (\times10^{-9})$ & $(0.0830, -0.0515, -0.0522, -0.0614)$ & $(0.0962, -0.0979)$ \\
		\hline
	\end{tabular}\label{tb: parameter for photon ring 2}
	}
\end{table}

\section{Conclusions and discussion}
\label{sec:conlude}

Horizonless ultracompact objects are a class of black hole mimickers that possess an unstable photon sphere.  The accretion disk around an ultracompact object can form outer photon rings under strong deflection gravitational lensing, similar to a black hole. But unlike the black hole, the ultracompact object can also have a stable antiphoton sphere, giving rise to distinctive inner photon rings. These inner rings are key to distinguishing the ultracompact object from the black hole.

In this paper, we present analytical descriptions of the shape, thickness and interferometric pattern of the inner and outer higher-order photon rings around an ultracompact object that has an unstable photon sphere but no event horizon. 
These descriptions are based on the strong deflection limit method with the finite distance effect for the black hole \cite{Bozza2007PRD76.083008} and for the ultracompact object \cite{Gao2024PRD109.063030}, providing a convenient way to explore the observational characteristics of the ultracompact objects in the radio wavelength band.

By taking the \SchStar with radius $R=2.5M$ as the gravitational lens and considering a thin circular emission disk on its equatorial plane, 
we find that
\begin{itemize}
    \item[(1)] The $n=2$ inner and outer photon rings are nearly circular for a low inclination.
    \item[(2)] The lower half of the $n=2$ inner ring can get clearly deformed for a high inclination.
    \item[(3)] The $n$th-order inner ring is always thicker than the $n$th-order outer ring.
    \item[(4)] The phase of the visibility is exactly determined by the tangential points on the shape curve of a photon ring.
    \item[(5)] The $n=2$ inner ring has more distinct features in both the intensity profile and the visibility than the $n=2$ outer ring, and can generate new envelopes in the visibility for a high inclination. These new envelopes have the potential to be detected by the Earth-Moon baseline interferometry.
    \item[(6)] By using the strong deflection limit approach, we can quickly and efficiently generate the interferometric pattern of the higher-order photon rings, which can match the numerical result well.
\end{itemize}

We do not consider the spin effect of the spacetime because the analytical method herein is established within the spherically symmetric framework, and numerical results have shown that such an effect on the shape and other properties of the photon rings should be small \cite{Vincent2022AAp667.A170, Paugnat2022AAp668.A11,EHTC2020PRL125.141104}.
Nevertheless, the spin of spacetime still holds profound theoretical and observational significance. For instance, the frame-dragging effect due to spin can induce significant azimuthal rotation within different levels of photon rings and echo signals \cite{Zhang2025arXiv2503.17200,Zhu2025arXiv2503.22343}, which may lead to new features in the corresponding interferometric signatures.
To include the spin effect, the key elements of analytical studies are the analytical metric describing a spinning ultracompact object \cite{Posada2017MNRAS468.2128,Hernandez-Pastora2017PRD95.024003,Ravi2018NA64.31,Kim2020PRD101.064067,Viaggiu2023IJMPD32.2350008} and the analytical approximation for light deflection in the spinning ultracompact object spacetime. However, the latter is still under development up to now \cite{Gralla2020PRD101.044031}, which means that constructing the shape curves of higher-order photon rings of the spinning ultracompact object can only rely on numerical algorithms for the time being \cite{Zhou2025PRD111.064075}.

Once the shape curves of the inner and outer photon rings around a spinning ultracompact object is obtained, we can still use the visibility approximation \eqref{Visibility: GAV} to generate the interferometric pattern since it is applicable to closed curves with arbitrary shape \cite{Gralla2020PRD102.044017}, and we expect that the introduction of the spin effect will not change its form, but only change the number of the tangential points.
However, according to the thickness analysis in Sec.~\ref{sec:chapter3}, it may be easy to go beyond the regime \eqref{regime_considered}, e.g., the baseline length $\mathrm{u}$ needs to be smaller than $50 \mathrm{G}\lambda$ (that roughly corresponds to the thickness of 1$M$) for the $n=2$ inner photon ring. Thus Eq.~\eqref{Visibility: GAV} should be generalized for $\mathrm{u}>50 \mathrm{G}\lambda$, which deserves future work.

The overall visibility of the outer photon rings is actually staircase-like due to their different intensities \cite{Johnson2020SciAdv6.eaaz1310,CardenasAvendano2023PRD108.064043}. 
The $n=1$ outer ring dominates the highest step, the $n=2$ outer ring dominates the second highest step and so on. 
The presence of the inner photon rings may change these relations.
Our analytical formulae can provide a computationally cheap way to study the correspondence between the visibility step height and the order of the photon ring. This will be our next move.

\appendix
\section{The projected intensity $P_x I_\mathrm{norm}$ and corresponding visibility of the $n=3$ photon rings}
\label{App：n3 photon ring}

\begin{figure}[t]    
	\centering
	\includegraphics[width=1.0\textwidth]{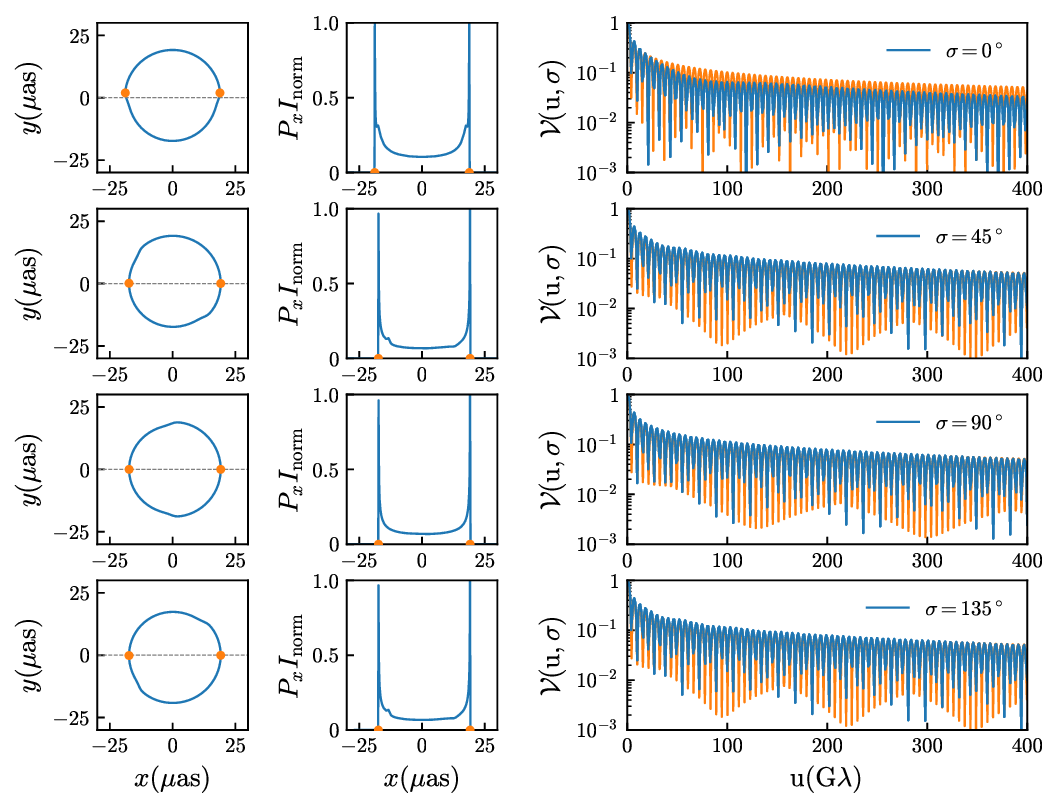}
	\caption{The shape curve, normalized projected intensity $P_x I_\mathrm{norm}$ and (normalized) visibility $\mathcal{V}(\mathrm{u}, \sigma)$ of the $n=3$ inner photon ring. The physical setup and color representation are the same as in Fig.~\ref{Fig4:PhotonRing_Inside_Visibility}.}
	\label{AppendixFig1:PhotonRing_Inside_Visibility}
\end{figure}

\begin{figure}[t]    
	\centering
	\includegraphics[width=1.0\textwidth]{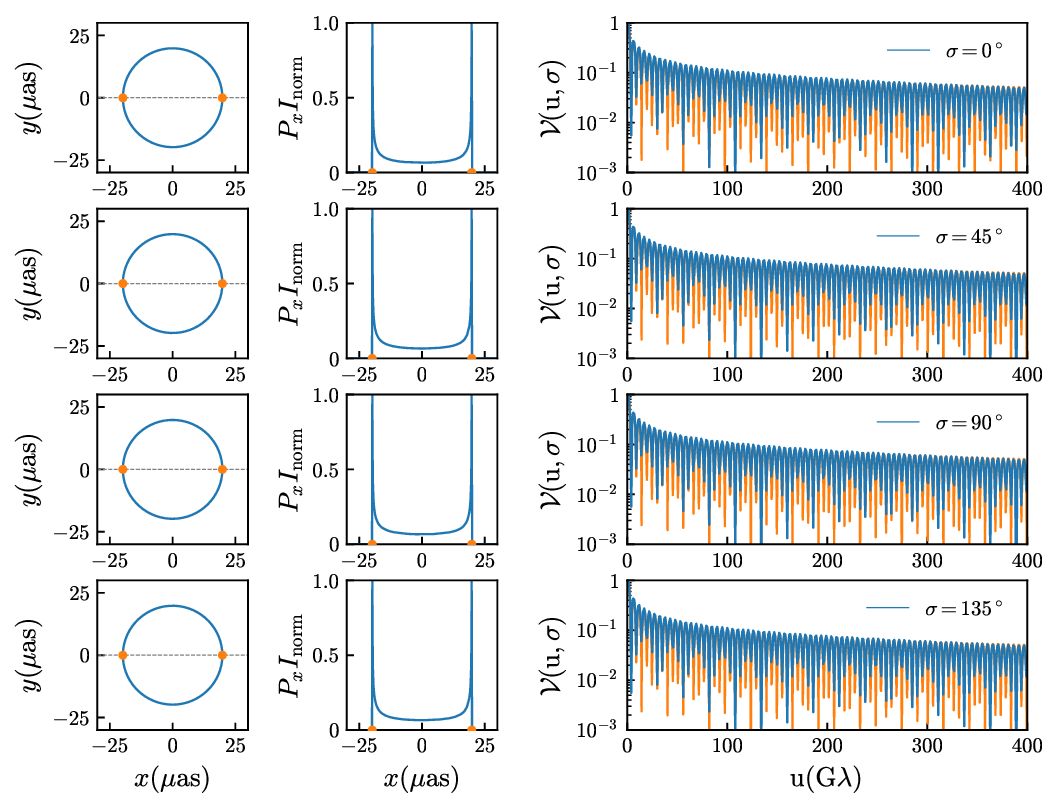}
	\caption{The shape curve, normalized projected intensity $P_x I_\mathrm{norm}$ and (normalized) visibility $\mathcal{V}(\mathrm{u}, \sigma)$ of the $n=3$ outer photon ring. The physical setup and color representation are the same as in Fig.~\ref{Fig4:PhotonRing_Inside_Visibility}.}
	\label{AppendixFig2:PhotonRing_Outside_Visibility}
\end{figure}

In Fig.~\ref{AppendixFig1:PhotonRing_Inside_Visibility}, we show the shape curve, normalized projected intensity $P_x I_\mathrm{norm}$ and (normalized) visibility $\mathcal{V}(\mathrm{u}, \sigma)$ of the $n=3$ inner ring.
In Fig.~\ref{AppendixFig2:PhotonRing_Outside_Visibility}, we show those of the $n=3$ outer ring.
In Table.~\ref{tb: parameter for photon ring 3}, $ \{ \mathcal{I}_\mathrm{T}, \mathcal{R}_\mathrm{T}, \mathcal{S}_\mathrm{T}, \mathcal{Z}_\mathrm{T} \} $ for the visibility of the $n=3$ photon rings around the Schwarzschild star with $R=2.5M$ are given.

\begin{table}[H]
	\centering   
	\caption{$ \{ \mathcal{I}_\mathrm{T}, \mathcal{R}_\mathrm{T}, \mathcal{S}_\mathrm{T}, \mathcal{Z}_\mathrm{T} \} $ for the visibility of the $n=3$ photon rings of the Schwarzschild star with radius $R=2.5M$.}
	\resizebox{0.8\textwidth}{!}{
	\begin{tabular}{llllcl}
		\hline
		& &\hspace{2.5cm} $n=3$ &  \\
		& & Inner ring & Outer ring \\
		\hline
		\multirow{4}{*}{$\sigma=0^\circ \phantom{\int XXX} $} 
		& $\mathcal{I}_\mathrm{T} \phantom{\int XXX}$ & $(1, 1)$ & $(1, 1)$ \\
		& $\mathcal{R}_\mathrm{T} (\times10^{-9}) \phantom{\int XXX}$ & $(0.0388, 0.0388)$ & $(0.0986, 0.0986)$ \\
		& $\mathcal{S}_\mathrm{T} \phantom{\int XXX}$ & $(-1, 1)$ & $(-1, 1)$ \\
		& $\mathcal{Z}_\mathrm{T} (\times10^{-9}) \phantom{\int XX}$ & $(0.0913, -0.0913)$ & $(0.0961, -0.0961)$ \\
		\hline
		\multirow{4}{*}{$\sigma=45^\circ$} 
		& $\mathcal{I}_\mathrm{T}$ & $(1, 1)$ & $(1, 1)$ \\
		& $\mathcal{R}_\mathrm{T} (\times10^{-9})$ & $(0.0923, 0.0861)$ & $(0.0961, 0.0961)$ \\
		& $\mathcal{S}_\mathrm{T}$ & $(-1, 1)$ & $(-1, 1)$ \\
		& $\mathcal{Z}_\mathrm{T} (\times10^{-9})$ & $(0.0928, -0.0842) \phantom{\int XXXX}$ & $(0.0960, -0.0961)$ \\
		\hline
		\multirow{4}{*}{$\sigma=90^\circ$} 
		& $\mathcal{I}_\mathrm{T}$ & $(1, 1)$ & $(1, 1)$ \\
		& $\mathcal{R}_\mathrm{T} (\times10^{-9})$ & $(0.0927, 0.0845)$ & $(0.0960, 0.0961)$ \\
		& $\mathcal{S}_\mathrm{T}$ & $(-1, 1)$ & $(1, 1)$ \\
		& $\mathcal{Z}_\mathrm{T} (\times10^{-9})$ & $(0.0929, -0.0841)$ & $(0.0960, -0.0961)$ \\
		\hline
		\multirow{4}{*}{$\sigma=135^\circ$} 
		& $\mathcal{I}_\mathrm{T}$ & $(1, 1)$ & $(1, 1)$ \\
		& $\mathcal{R}_\mathrm{T} (\times10^{-9})$ & $(0.0923, 0.0861)$ & $(0.0961, 0.0961)$ \\
		& $\mathcal{S}_\mathrm{T}$ & $(-1,1)$ & $(-1, 1)$ \\
		& $\mathcal{Z}_\mathrm{T} (\times10^{-9})$ & $(0.0928, -0.0842)$ & $(0.0960, -0.0961)$ \\
		\hline
	\end{tabular}\label{tb: parameter for photon ring 3}
	}
\end{table}

\section*{Acknowledgements}
We thank Prof. Yi Xie for helpful discussions.
This work is funded by the National Natural Science Foundation of China (Grants Nos. 12447143, 12273116, 62394350 and 62394351), the science research grants from the China Manned Space Project (Grants Nos. CMS-CSST-2021-A12 and CMS-CSST-2021-B10), the Strategic Priority Research Program of the Chinese Academy of Sciences (Grant No.XDA0350302) and the Opening Project of National Key Laboratory of Aerospace Flight Dynamics of China (Grant No. KGJ6142210220201).

\bibliographystyle{apsrev4-1} 
\bibliography{Refs20230310} 

\begin{thebibliography}{83}%
\makeatletter
\providecommand \@ifxundefined [1]{%
 \@ifx{#1\undefined}
}%
\providecommand \@ifnum [1]{%
 \ifnum #1\expandafter \@firstoftwo
 \else \expandafter \@secondoftwo
 \fi
}%
\providecommand \@ifx [1]{%
 \ifx #1\expandafter \@firstoftwo
 \else \expandafter \@secondoftwo
 \fi
}%
\providecommand \natexlab [1]{#1}%
\providecommand \enquote  [1]{``#1''}%
\providecommand \bibnamefont  [1]{#1}%
\providecommand \bibfnamefont [1]{#1}%
\providecommand \citenamefont [1]{#1}%
\providecommand \href@noop [0]{\@secondoftwo}%
\providecommand \href [0]{\begingroup \@sanitize@url \@href}%
\providecommand \@href[1]{\@@startlink{#1}\@@href}%
\providecommand \@@href[1]{\endgroup#1\@@endlink}%
\providecommand \@sanitize@url [0]{\catcode `\\12\catcode `\$12\catcode
  `\&12\catcode `\#12\catcode `\^12\catcode `\_12\catcode `\%12\relax}%
\providecommand \@@startlink[1]{}%
\providecommand \@@endlink[0]{}%
\providecommand \url  [0]{\begingroup\@sanitize@url \@url }%
\providecommand \@url [1]{\endgroup\@href {#1}{\urlprefix }}%
\providecommand \urlprefix  [0]{URL }%
\providecommand \Eprint [0]{\href }%
\providecommand \doibase [0]{http://dx.doi.org/}%
\providecommand \selectlanguage [0]{\@gobble}%
\providecommand \bibinfo  [0]{\@secondoftwo}%
\providecommand \bibfield  [0]{\@secondoftwo}%
\providecommand \translation [1]{[#1]}%
\providecommand \BibitemOpen [0]{}%
\providecommand \bibitemStop [0]{}%
\providecommand \bibitemNoStop [0]{.\EOS\space}%
\providecommand \EOS [0]{\spacefactor3000\relax}%
\providecommand \BibitemShut  [1]{\csname bibitem#1\endcsname}%
\let\auto@bib@innerbib\@empty
\bibitem [{\citenamefont {{ B.~P. {Abbott \textit{et al.} (LIGO Scientific and
  Virgo Collaborations)}}}(2016{\natexlab{a}})}]{LVC2016PRL116.061102}%
  \BibitemOpen
  \bibfield  {author} {\bibinfo {author} {\bibnamefont {{ B.~P. {Abbott
  \textit{et al.} (LIGO Scientific and Virgo Collaborations)}}}},\ }\href
  {\doibase 10.1103/PhysRevLett.116.061102} {\bibfield  {journal} {\bibinfo
  {journal} {\prl}\ }\textbf {\bibinfo {volume} {116}},\ \bibinfo {eid}
  {061102} (\bibinfo {year} {2016}{\natexlab{a}})}\BibitemShut {NoStop}%
\bibitem [{\citenamefont {{ B.~P. {Abbott \textit{et al.} (LIGO Scientific and
  Virgo Collaborations)}}}(2016{\natexlab{b}})}]{LVC2016PRX6.041015}%
  \BibitemOpen
  \bibfield  {author} {\bibinfo {author} {\bibnamefont {{ B.~P. {Abbott
  \textit{et al.} (LIGO Scientific and Virgo Collaborations)}}}},\ }\href
  {\doibase 10.1103/PhysRevX.6.041015} {\bibfield  {journal} {\bibinfo
  {journal} {\prx}\ }\textbf {\bibinfo {volume} {6}},\ \bibinfo {eid} {041015}
  (\bibinfo {year} {2016}{\natexlab{b}})}\BibitemShut {NoStop}%
\bibitem [{\citenamefont {{ B.~P. {Abbott \textit{et al.} (LIGO Scientific and
  Virgo Collaborations)}}}(2016{\natexlab{c}})}]{LVC2016PRL116.241103}%
  \BibitemOpen
  \bibfield  {author} {\bibinfo {author} {\bibnamefont {{ B.~P. {Abbott
  \textit{et al.} (LIGO Scientific and Virgo Collaborations)}}}},\ }\href
  {\doibase 10.1103/PhysRevLett.116.241103} {\bibfield  {journal} {\bibinfo
  {journal} {\prl}\ }\textbf {\bibinfo {volume} {116}},\ \bibinfo {eid}
  {241103} (\bibinfo {year} {2016}{\natexlab{c}})}\BibitemShut {NoStop}%
\bibitem [{\citenamefont {{ B.~P. {Abbott \textit{et al.} (LIGO Scientific and
  Virgo Collaborations)}}}(2017{\natexlab{a}})}]{LVC2017PRL118.221101}%
  \BibitemOpen
  \bibfield  {author} {\bibinfo {author} {\bibnamefont {{ B.~P. {Abbott
  \textit{et al.} (LIGO Scientific and Virgo Collaborations)}}}},\ }\href
  {\doibase 10.1103/PhysRevLett.118.221101} {\bibfield  {journal} {\bibinfo
  {journal} {\prl}\ }\textbf {\bibinfo {volume} {118}},\ \bibinfo {eid}
  {221101} (\bibinfo {year} {2017}{\natexlab{a}})}\BibitemShut {NoStop}%
\bibitem [{\citenamefont {{ B.~P. {Abbott \textit{et al.} (LIGO Scientific and
  Virgo Collaborations)}}}(2017{\natexlab{b}})}]{LVC2017ApJ851.L35}%
  \BibitemOpen
  \bibfield  {author} {\bibinfo {author} {\bibnamefont {{ B.~P. {Abbott
  \textit{et al.} (LIGO Scientific and Virgo Collaborations)}}}},\ }\href
  {\doibase 10.3847/2041-8213/aa9f0c} {\bibfield  {journal} {\bibinfo
  {journal} {\apjl}\ }\textbf {\bibinfo {volume} {851}},\ \bibinfo {eid} {L35}
  (\bibinfo {year} {2017}{\natexlab{b}})}\BibitemShut {NoStop}%
\bibitem [{\citenamefont {{ B.~P. {Abbott \textit{et al.} (LIGO Scientific and
  Virgo Collaborations)}}}(2017{\natexlab{c}})}]{LVC2017PRL119.141101}%
  \BibitemOpen
  \bibfield  {author} {\bibinfo {author} {\bibnamefont {{ B.~P. {Abbott
  \textit{et al.} (LIGO Scientific and Virgo Collaborations)}}}},\ }\href
  {\doibase 10.1103/PhysRevLett.119.141101} {\bibfield  {journal} {\bibinfo
  {journal} {\prl}\ }\textbf {\bibinfo {volume} {119}},\ \bibinfo {eid}
  {141101} (\bibinfo {year} {2017}{\natexlab{c}})}\BibitemShut {NoStop}%
\bibitem [{\citenamefont {{Akiyama \textit{et al.} (Event Horizon Telescope
  Collaboration)}}(2019{\natexlab{a}})}]{EHTC2019ApJ875.L1}%
  \BibitemOpen
  \bibfield  {author} {\bibinfo {author} {\bibfnamefont {K.}~\bibnamefont
  {{Akiyama \textit{et al.} (Event Horizon Telescope Collaboration)}}},\ }\href
  {\doibase 10.3847/2041-8213/ab0ec7} {\bibfield  {journal} {\bibinfo
  {journal} {\apjl}\ }\textbf {\bibinfo {volume} {875}},\ \bibinfo {eid} {L1}
  (\bibinfo {year} {2019}{\natexlab{a}})}\BibitemShut {NoStop}%
\bibitem [{\citenamefont {{Akiyama \textit{et al.} (Event Horizon Telescope
  Collaboration)}}(2019{\natexlab{b}})}]{EHTC2019ApJ875.L2}%
  \BibitemOpen
  \bibfield  {author} {\bibinfo {author} {\bibfnamefont {K.}~\bibnamefont
  {{Akiyama \textit{et al.} (Event Horizon Telescope Collaboration)}}},\ }\href
  {\doibase 10.3847/2041-8213/ab0c96} {\bibfield  {journal} {\bibinfo
  {journal} {\apjl}\ }\textbf {\bibinfo {volume} {875}},\ \bibinfo {eid} {L2}
  (\bibinfo {year} {2019}{\natexlab{b}})}\BibitemShut {NoStop}%
\bibitem [{\citenamefont {{Akiyama \textit{et al.} (Event Horizon Telescope
  Collaboration)}}(2019{\natexlab{c}})}]{EHTC2019ApJ875.L3}%
  \BibitemOpen
  \bibfield  {author} {\bibinfo {author} {\bibfnamefont {K.}~\bibnamefont
  {{Akiyama \textit{et al.} (Event Horizon Telescope Collaboration)}}},\ }\href
  {\doibase 10.3847/2041-8213/ab0c57} {\bibfield  {journal} {\bibinfo
  {journal} {\apjl}\ }\textbf {\bibinfo {volume} {875}},\ \bibinfo {eid} {L3}
  (\bibinfo {year} {2019}{\natexlab{c}})}\BibitemShut {NoStop}%
\bibitem [{\citenamefont {{Akiyama \textit{et al.} (Event Horizon Telescope
  Collaboration)}}(2019{\natexlab{d}})}]{EHTC2019ApJ875.L4}%
  \BibitemOpen
  \bibfield  {author} {\bibinfo {author} {\bibfnamefont {K.}~\bibnamefont
  {{Akiyama \textit{et al.} (Event Horizon Telescope Collaboration)}}},\ }\href
  {\doibase 10.3847/2041-8213/ab0e85} {\bibfield  {journal} {\bibinfo
  {journal} {\apjl}\ }\textbf {\bibinfo {volume} {875}},\ \bibinfo {eid} {L4}
  (\bibinfo {year} {2019}{\natexlab{d}})}\BibitemShut {NoStop}%
\bibitem [{\citenamefont {{Akiyama \textit{et al.} (Event Horizon Telescope
  Collaboration)}}(2019{\natexlab{e}})}]{EHTC2019ApJ875.L5}%
  \BibitemOpen
  \bibfield  {author} {\bibinfo {author} {\bibfnamefont {K.}~\bibnamefont
  {{Akiyama \textit{et al.} (Event Horizon Telescope Collaboration)}}},\ }\href
  {\doibase 10.3847/2041-8213/ab0f43} {\bibfield  {journal} {\bibinfo
  {journal} {\apjl}\ }\textbf {\bibinfo {volume} {875}},\ \bibinfo {eid} {L5}
  (\bibinfo {year} {2019}{\natexlab{e}})}\BibitemShut {NoStop}%
\bibitem [{\citenamefont {{Akiyama \textit{et al.} (Event Horizon Telescope
  Collaboration)}}(2019{\natexlab{f}})}]{EHTC2019ApJ875.L6}%
  \BibitemOpen
  \bibfield  {author} {\bibinfo {author} {\bibfnamefont {K.}~\bibnamefont
  {{Akiyama \textit{et al.} (Event Horizon Telescope Collaboration)}}},\ }\href
  {\doibase 10.3847/2041-8213/ab1141} {\bibfield  {journal} {\bibinfo
  {journal} {\apjl}\ }\textbf {\bibinfo {volume} {875}},\ \bibinfo {eid} {L6}
  (\bibinfo {year} {2019}{\natexlab{f}})}\BibitemShut {NoStop}%
\bibitem [{\citenamefont {{Akiyama \textit{et al.} (Event Horizon Telescope
  Collaboration)}}(2022{\natexlab{a}})}]{EHTC2022ApJ930.L12}%
  \BibitemOpen
  \bibfield  {author} {\bibinfo {author} {\bibfnamefont {K.}~\bibnamefont
  {{Akiyama \textit{et al.} (Event Horizon Telescope Collaboration)}}},\ }\href
  {\doibase 10.3847/2041-8213/ac6674} {\bibfield  {journal} {\bibinfo
  {journal} {\apjl}\ }\textbf {\bibinfo {volume} {930}},\ \bibinfo {eid} {L12}
  (\bibinfo {year} {2022}{\natexlab{a}})}\BibitemShut {NoStop}%
\bibitem [{\citenamefont {{Akiyama \textit{et al.} (Event Horizon Telescope
  Collaboration)}}(2022{\natexlab{b}})}]{EHTC2022ApJ930.L13}%
  \BibitemOpen
  \bibfield  {author} {\bibinfo {author} {\bibfnamefont {K.}~\bibnamefont
  {{Akiyama \textit{et al.} (Event Horizon Telescope Collaboration)}}},\ }\href
  {\doibase 10.3847/2041-8213/ac6675} {\bibfield  {journal} {\bibinfo
  {journal} {\apjl}\ }\textbf {\bibinfo {volume} {930}},\ \bibinfo {eid} {L13}
  (\bibinfo {year} {2022}{\natexlab{b}})}\BibitemShut {NoStop}%
\bibitem [{\citenamefont {{Akiyama \textit{et al.} (Event Horizon Telescope
  Collaboration)}}(2022{\natexlab{c}})}]{EHTC2022ApJ930.L14}%
  \BibitemOpen
  \bibfield  {author} {\bibinfo {author} {\bibfnamefont {K.}~\bibnamefont
  {{Akiyama \textit{et al.} (Event Horizon Telescope Collaboration)}}},\ }\href
  {\doibase 10.3847/2041-8213/ac6429} {\bibfield  {journal} {\bibinfo
  {journal} {\apjl}\ }\textbf {\bibinfo {volume} {930}},\ \bibinfo {eid} {L14}
  (\bibinfo {year} {2022}{\natexlab{c}})}\BibitemShut {NoStop}%
\bibitem [{\citenamefont {{Akiyama \textit{et al.} (Event Horizon Telescope
  Collaboration)}}(2022{\natexlab{d}})}]{EHTC2022ApJ930.L15}%
  \BibitemOpen
  \bibfield  {author} {\bibinfo {author} {\bibfnamefont {K.}~\bibnamefont
  {{Akiyama \textit{et al.} (Event Horizon Telescope Collaboration)}}},\ }\href
  {\doibase 10.3847/2041-8213/ac6736} {\bibfield  {journal} {\bibinfo
  {journal} {\apjl}\ }\textbf {\bibinfo {volume} {930}},\ \bibinfo {eid} {L15}
  (\bibinfo {year} {2022}{\natexlab{d}})}\BibitemShut {NoStop}%
\bibitem [{\citenamefont {{Akiyama \textit{et al.} (Event Horizon Telescope
  Collaboration)}}(2022{\natexlab{e}})}]{EHTC2022ApJ930.L16}%
  \BibitemOpen
  \bibfield  {author} {\bibinfo {author} {\bibfnamefont {K.}~\bibnamefont
  {{Akiyama \textit{et al.} (Event Horizon Telescope Collaboration)}}},\ }\href
  {\doibase 10.3847/2041-8213/ac6672} {\bibfield  {journal} {\bibinfo
  {journal} {\apjl}\ }\textbf {\bibinfo {volume} {930}},\ \bibinfo {eid} {L16}
  (\bibinfo {year} {2022}{\natexlab{e}})}\BibitemShut {NoStop}%
\bibitem [{\citenamefont {{Akiyama \textit{et al.} (Event Horizon Telescope
  Collaboration)}}(2022{\natexlab{f}})}]{EHTC2022ApJ930.L17}%
  \BibitemOpen
  \bibfield  {author} {\bibinfo {author} {\bibfnamefont {K.}~\bibnamefont
  {{Akiyama \textit{et al.} (Event Horizon Telescope Collaboration)}}},\ }\href
  {\doibase 10.3847/2041-8213/ac6756} {\bibfield  {journal} {\bibinfo
  {journal} {\apjl}\ }\textbf {\bibinfo {volume} {930}},\ \bibinfo {eid} {L17}
  (\bibinfo {year} {2022}{\natexlab{f}})}\BibitemShut {NoStop}%
\bibitem [{\citenamefont {Hawking}(1975)}]{Hawking1975CMP43.199}%
  \BibitemOpen
  \bibfield  {author} {\bibinfo {author} {\bibfnamefont {S.~W.}\ \bibnamefont
  {Hawking}},\ }\href {\doibase 10.1007/BF02345020} {\bibfield  {journal}
  {\bibinfo  {journal} {Commun. Math. Phys.}\ }\textbf {\bibinfo {volume}
  {43}},\ \bibinfo {pages} {199} (\bibinfo {year} {1975})},\ \bibinfo {note}
  {[Erratum: Commun.Math.Phys. 46, 206 (1976)]}\BibitemShut {NoStop}%
\bibitem [{\citenamefont {{Mathur}}(2009)}]{Mathur2009CQG26.224001}%
  \BibitemOpen
  \bibfield  {author} {\bibinfo {author} {\bibfnamefont {S.~D.}\ \bibnamefont
  {{Mathur}}},\ }\href {\doibase 10.1088/0264-9381/26/22/224001} {\bibfield
  {journal} {\bibinfo  {journal} {\cqg}\ }\textbf {\bibinfo {volume} {26}},\
  \bibinfo {eid} {224001} (\bibinfo {year} {2009})}\BibitemShut {NoStop}%
\bibitem [{\citenamefont {{Cardoso}}\ and\ \citenamefont
  {{Pani}}(2019)}]{Cardoso2019LRR22.4}%
  \BibitemOpen
  \bibfield  {author} {\bibinfo {author} {\bibfnamefont {V.}~\bibnamefont
  {{Cardoso}}}\ and\ \bibinfo {author} {\bibfnamefont {P.}~\bibnamefont
  {{Pani}}},\ }\href {\doibase 10.1007/s41114-019-0020-4} {\bibfield  {journal}
  {\bibinfo  {journal} {\lrr}\ }\textbf {\bibinfo {volume} {22}},\ \bibinfo
  {eid} {4} (\bibinfo {year} {2019})}\BibitemShut {NoStop}%
\bibitem [{\citenamefont {{Schwarzschild}}(1916)}]{Schwarzschild1916SPAWB424}%
  \BibitemOpen
  \bibfield  {author} {\bibinfo {author} {\bibfnamefont {K.}~\bibnamefont
  {{Schwarzschild}}},\ }\href@noop {} {\bibfield  {journal} {\bibinfo
  {journal} {Sitzungsber. Preuss. Akad. Wiss. Berlin (Math. Phys. )}\ }\textbf
  {\bibinfo {volume} {1916}},\ \bibinfo {pages} {424} (\bibinfo {year}
  {1916})}\BibitemShut {NoStop}%
\bibitem [{\citenamefont {{Wald}}(1984)}]{Wald1984Book}%
  \BibitemOpen
  \bibfield  {author} {\bibinfo {author} {\bibfnamefont {R.~M.}\ \bibnamefont
  {{Wald}}},\ }\href@noop {} {\emph {\bibinfo {title} {{General Relativity}}}}\
  (\bibinfo  {publisher} {University of Chicago Press},\ \bibinfo {address}
  {Chicago},\ \bibinfo {year} {1984})\BibitemShut {NoStop}%
\bibitem [{\citenamefont {{Cardoso}}\ and\ \citenamefont
  {{Pani}}(2017)}]{Cardoso2017NatAst1.586}%
  \BibitemOpen
  \bibfield  {author} {\bibinfo {author} {\bibfnamefont {V.}~\bibnamefont
  {{Cardoso}}}\ and\ \bibinfo {author} {\bibfnamefont {P.}~\bibnamefont
  {{Pani}}},\ }\href {\doibase 10.1038/s41550-017-0225-y} {\bibfield  {journal}
  {\bibinfo  {journal} {\natast}\ }\textbf {\bibinfo {volume} {1}},\ \bibinfo
  {pages} {586} (\bibinfo {year} {2017})}\BibitemShut {NoStop}%
\bibitem [{\citenamefont {{Berti}}\ \emph {et~al.}(2016)\citenamefont
  {{Berti}}, \citenamefont {{Sesana}}, \citenamefont {{Barausse}},
  \citenamefont {{Cardoso}},\ and\ \citenamefont
  {{Belczynski}}}]{Berti2016PRL117.101102}%
  \BibitemOpen
  \bibfield  {author} {\bibinfo {author} {\bibfnamefont {E.}~\bibnamefont
  {{Berti}}}, \bibinfo {author} {\bibfnamefont {A.}~\bibnamefont {{Sesana}}},
  \bibinfo {author} {\bibfnamefont {E.}~\bibnamefont {{Barausse}}}, \bibinfo
  {author} {\bibfnamefont {V.}~\bibnamefont {{Cardoso}}}, \ and\ \bibinfo
  {author} {\bibfnamefont {K.}~\bibnamefont {{Belczynski}}},\ }\href {\doibase
  10.1103/PhysRevLett.117.101102} {\bibfield  {journal} {\bibinfo  {journal}
  {\prl}\ }\textbf {\bibinfo {volume} {117}},\ \bibinfo {eid} {101102}
  (\bibinfo {year} {2016})}\BibitemShut {NoStop}%
\bibitem [{\citenamefont {{Gralla}}\ \emph {et~al.}(2019)\citenamefont
  {{Gralla}}, \citenamefont {{Holz}},\ and\ \citenamefont
  {{Wald}}}]{Gralla2019PRD100.024018}%
  \BibitemOpen
  \bibfield  {author} {\bibinfo {author} {\bibfnamefont {S.~E.}\ \bibnamefont
  {{Gralla}}}, \bibinfo {author} {\bibfnamefont {D.~E.}\ \bibnamefont
  {{Holz}}}, \ and\ \bibinfo {author} {\bibfnamefont {R.~M.}\ \bibnamefont
  {{Wald}}},\ }\href {\doibase 10.1103/PhysRevD.100.024018} {\bibfield
  {journal} {\bibinfo  {journal} {\prd}\ }\textbf {\bibinfo {volume} {100}},\
  \bibinfo {eid} {024018} (\bibinfo {year} {2019})}\BibitemShut {NoStop}%
\bibitem [{\citenamefont {{Johnson}}\ \emph {et~al.}(2020)\citenamefont
  {{Johnson}}, \citenamefont {{Lupsasca}}, \citenamefont {{Strominger}},
  \citenamefont {{Wong}}, \citenamefont {{Hadar}}, \citenamefont {{Kapec}},
  \citenamefont {{Narayan}}, \citenamefont {{Chael}}, \citenamefont {{Gammie}},
  \citenamefont {{Galison}}, \citenamefont {{Palumbo}}, \citenamefont
  {{Doeleman}}, \citenamefont {{Blackburn}}, \citenamefont {{Wielgus}},
  \citenamefont {{Pesce}}, \citenamefont {{Farah}},\ and\ \citenamefont
  {{Moran}}}]{Johnson2020SciAdv6.eaaz1310}%
  \BibitemOpen
  \bibfield  {author} {\bibinfo {author} {\bibfnamefont {M.~D.}\ \bibnamefont
  {{Johnson}}}, \bibinfo {author} {\bibfnamefont {A.}~\bibnamefont
  {{Lupsasca}}}, \bibinfo {author} {\bibfnamefont {A.}~\bibnamefont
  {{Strominger}}}, \bibinfo {author} {\bibfnamefont {G.~N.}\ \bibnamefont
  {{Wong}}}, \bibinfo {author} {\bibfnamefont {S.}~\bibnamefont {{Hadar}}},
  \bibinfo {author} {\bibfnamefont {D.}~\bibnamefont {{Kapec}}}, \bibinfo
  {author} {\bibfnamefont {R.}~\bibnamefont {{Narayan}}}, \bibinfo {author}
  {\bibfnamefont {A.}~\bibnamefont {{Chael}}}, \bibinfo {author} {\bibfnamefont
  {C.~F.}\ \bibnamefont {{Gammie}}}, \bibinfo {author} {\bibfnamefont
  {P.}~\bibnamefont {{Galison}}}, \bibinfo {author} {\bibfnamefont {D.~C.~M.}\
  \bibnamefont {{Palumbo}}}, \bibinfo {author} {\bibfnamefont {S.~S.}\
  \bibnamefont {{Doeleman}}}, \bibinfo {author} {\bibfnamefont
  {L.}~\bibnamefont {{Blackburn}}}, \bibinfo {author} {\bibfnamefont
  {M.}~\bibnamefont {{Wielgus}}}, \bibinfo {author} {\bibfnamefont {D.~W.}\
  \bibnamefont {{Pesce}}}, \bibinfo {author} {\bibfnamefont {J.~R.}\
  \bibnamefont {{Farah}}}, \ and\ \bibinfo {author} {\bibfnamefont {J.~M.}\
  \bibnamefont {{Moran}}},\ }\href {\doibase 10.1126/sciadv.aaz1310} {\bibfield
   {journal} {\bibinfo  {journal} {\sciadv}\ }\textbf {\bibinfo {volume} {6}},\
  \bibinfo {pages} {eaaz1310} (\bibinfo {year} {2020})}\BibitemShut {NoStop}%
\bibitem [{\citenamefont {{Vincent}}\ \emph {et~al.}(2022)\citenamefont
  {{Vincent}}, \citenamefont {{Gralla}}, \citenamefont {{Lupsasca}},\ and\
  \citenamefont {{Wielgus}}}]{Vincent2022AAp667.A170}%
  \BibitemOpen
  \bibfield  {author} {\bibinfo {author} {\bibfnamefont {F.~H.}\ \bibnamefont
  {{Vincent}}}, \bibinfo {author} {\bibfnamefont {S.~E.}\ \bibnamefont
  {{Gralla}}}, \bibinfo {author} {\bibfnamefont {A.}~\bibnamefont
  {{Lupsasca}}}, \ and\ \bibinfo {author} {\bibfnamefont {M.}~\bibnamefont
  {{Wielgus}}},\ }\href {\doibase 10.1051/0004-6361/202244339} {\bibfield
  {journal} {\bibinfo  {journal} {\aap}\ }\textbf {\bibinfo {volume} {667}},\
  \bibinfo {pages} {A170} (\bibinfo {year} {2022})}\BibitemShut {NoStop}%
\bibitem [{\citenamefont {{Paugnat}}\ \emph {et~al.}(2022)\citenamefont
  {{Paugnat}}, \citenamefont {{Lupsasca}}, \citenamefont {{Vincent}},\ and\
  \citenamefont {{Wielgus}}}]{Paugnat2022AAp668.A11}%
  \BibitemOpen
  \bibfield  {author} {\bibinfo {author} {\bibfnamefont {H.}~\bibnamefont
  {{Paugnat}}}, \bibinfo {author} {\bibfnamefont {A.}~\bibnamefont
  {{Lupsasca}}}, \bibinfo {author} {\bibfnamefont {F.~H.}\ \bibnamefont
  {{Vincent}}}, \ and\ \bibinfo {author} {\bibfnamefont {M.}~\bibnamefont
  {{Wielgus}}},\ }\href {\doibase 10.1051/0004-6361/202244216} {\bibfield
  {journal} {\bibinfo  {journal} {\aap}\ }\textbf {\bibinfo {volume} {668}},\
  \bibinfo {pages} {A11} (\bibinfo {year} {2022})}\BibitemShut {NoStop}%
\bibitem [{\citenamefont {{Virbhadra}}\ and\ \citenamefont
  {{Ellis}}(2000)}]{Virbhadra2000PRD62.084003}%
  \BibitemOpen
  \bibfield  {author} {\bibinfo {author} {\bibfnamefont {K.~S.}\ \bibnamefont
  {{Virbhadra}}}\ and\ \bibinfo {author} {\bibfnamefont {G.~F.~R.}\
  \bibnamefont {{Ellis}}},\ }\href@noop {} {\bibfield  {journal} {\bibinfo
  {journal} {\prd}\ }\textbf {\bibinfo {volume} {62}},\ \bibinfo {eid} {084003}
  (\bibinfo {year} {2000})}\BibitemShut {NoStop}%
\bibitem [{\citenamefont {{Bozza}}\ \emph {et~al.}(2001)\citenamefont
  {{Bozza}}, \citenamefont {{Capozziello}}, \citenamefont {{Iovane}},\ and\
  \citenamefont {{Scarpetta}}}]{Bozza2001GRG33.1535}%
  \BibitemOpen
  \bibfield  {author} {\bibinfo {author} {\bibfnamefont {V.}~\bibnamefont
  {{Bozza}}}, \bibinfo {author} {\bibfnamefont {S.}~\bibnamefont
  {{Capozziello}}}, \bibinfo {author} {\bibfnamefont {G.}~\bibnamefont
  {{Iovane}}}, \ and\ \bibinfo {author} {\bibfnamefont {G.}~\bibnamefont
  {{Scarpetta}}},\ }\href {\doibase 10.1023/A:1012292927358} {\bibfield
  {journal} {\bibinfo  {journal} {\grg}\ }\textbf {\bibinfo {volume} {33}},\
  \bibinfo {pages} {1535} (\bibinfo {year} {2001})}\BibitemShut {NoStop}%
\bibitem [{\citenamefont {{Bozza}}\ and\ \citenamefont
  {{Scarpetta}}(2007)}]{Bozza2007PRD76.083008}%
  \BibitemOpen
  \bibfield  {author} {\bibinfo {author} {\bibfnamefont {V.}~\bibnamefont
  {{Bozza}}}\ and\ \bibinfo {author} {\bibfnamefont {G.}~\bibnamefont
  {{Scarpetta}}},\ }\href {\doibase 10.1103/PhysRevD.76.083008} {\bibfield
  {journal} {\bibinfo  {journal} {\prd}\ }\textbf {\bibinfo {volume} {76}},\
  \bibinfo {eid} {083008} (\bibinfo {year} {2007})}\BibitemShut {NoStop}%
\bibitem [{\citenamefont {{Shaikh}}\ \emph {et~al.}(2019)\citenamefont
  {{Shaikh}}, \citenamefont {{Banerjee}}, \citenamefont {{Paul}},\ and\
  \citenamefont {{Sarkar}}}]{Shaikh2019PRD99.104040}%
  \BibitemOpen
  \bibfield  {author} {\bibinfo {author} {\bibfnamefont {R.}~\bibnamefont
  {{Shaikh}}}, \bibinfo {author} {\bibfnamefont {P.}~\bibnamefont
  {{Banerjee}}}, \bibinfo {author} {\bibfnamefont {S.}~\bibnamefont {{Paul}}},
  \ and\ \bibinfo {author} {\bibfnamefont {T.}~\bibnamefont {{Sarkar}}},\
  }\href {\doibase 10.1103/PhysRevD.99.104040} {\bibfield  {journal} {\bibinfo
  {journal} {\prd}\ }\textbf {\bibinfo {volume} {99}},\ \bibinfo {eid} {104040}
  (\bibinfo {year} {2019})}\BibitemShut {NoStop}%
\bibitem [{\citenamefont {{Petters}}(2003)}]{Petters2003MNRAS338.457}%
  \BibitemOpen
  \bibfield  {author} {\bibinfo {author} {\bibfnamefont {A.~O.}\ \bibnamefont
  {{Petters}}},\ }\href {\doibase 10.1046/j.1365-8711.2003.06065.x} {\bibfield
  {journal} {\bibinfo  {journal} {\mnras}\ }\textbf {\bibinfo {volume} {338}},\
  \bibinfo {pages} {457} (\bibinfo {year} {2003})}\BibitemShut {NoStop}%
\bibitem [{\citenamefont {{Aratore}}\ and\ \citenamefont
  {{Bozza}}(2021)}]{Aratore2021JCAP10.054}%
  \BibitemOpen
  \bibfield  {author} {\bibinfo {author} {\bibfnamefont {F.}~\bibnamefont
  {{Aratore}}}\ and\ \bibinfo {author} {\bibfnamefont {V.}~\bibnamefont
  {{Bozza}}},\ }\href {\doibase 10.1088/1475-7516/2021/10/054} {\bibfield
  {journal} {\bibinfo  {journal} {\jcap}\ }\textbf {\bibinfo {volume} {10}},\
  \bibinfo {pages} {054} (\bibinfo {year} {2021})}\BibitemShut {NoStop}%
\bibitem [{\citenamefont {{Gao}}\ and\ \citenamefont
  {{Xie}}(2022)}]{Gao2022EPJC82.162}%
  \BibitemOpen
  \bibfield  {author} {\bibinfo {author} {\bibfnamefont {Y.-X.}\ \bibnamefont
  {{Gao}}}\ and\ \bibinfo {author} {\bibfnamefont {Y.}~\bibnamefont {{Xie}}},\
  }\href {\doibase 10.1140/epjc/s10052-022-10128-z} {\bibfield  {journal}
  {\bibinfo  {journal} {\epjc}\ }\textbf {\bibinfo {volume} {82}},\ \bibinfo
  {eid} {162} (\bibinfo {year} {2022})}\BibitemShut {NoStop}%
\bibitem [{\citenamefont {Gao}\ and\ \citenamefont
  {Xie}(2024)}]{Gao2024PRD109.063030}%
  \BibitemOpen
  \bibfield  {author} {\bibinfo {author} {\bibfnamefont {Y.-X.}\ \bibnamefont
  {Gao}}\ and\ \bibinfo {author} {\bibfnamefont {Y.}~\bibnamefont {Xie}},\
  }\href {\doibase 10.1103/PhysRevD.109.063030} {\bibfield  {journal} {\bibinfo
   {journal} {\prd}\ }\textbf {\bibinfo {volume} {109}},\ \bibinfo {pages}
  {063030} (\bibinfo {year} {2024})}\BibitemShut {NoStop}%
\bibitem [{\citenamefont {{Chen}}\ \emph {et~al.}(2024)\citenamefont {{Chen}},
  \citenamefont {{Wang}}, \citenamefont {{Wu}},\ and\ \citenamefont
  {{Yang}}}]{Chen2024JCAP04.032}%
  \BibitemOpen
  \bibfield  {author} {\bibinfo {author} {\bibfnamefont {Y.}~\bibnamefont
  {{Chen}}}, \bibinfo {author} {\bibfnamefont {P.}~\bibnamefont {{Wang}}},
  \bibinfo {author} {\bibfnamefont {H.}~\bibnamefont {{Wu}}}, \ and\ \bibinfo
  {author} {\bibfnamefont {H.}~\bibnamefont {{Yang}}},\ }\href {\doibase
  10.1088/1475-7516/2024/04/032} {\bibfield  {journal} {\bibinfo  {journal}
  {\jcap}\ }\textbf {\bibinfo {volume} {2024}},\ \bibinfo {eid} {032} (\bibinfo
  {year} {2024})}\BibitemShut {NoStop}%
\bibitem [{\citenamefont {{Tsupko}}(2022)}]{Tsupko2022PRD106.064033}%
  \BibitemOpen
  \bibfield  {author} {\bibinfo {author} {\bibfnamefont {O.~Y.}\ \bibnamefont
  {{Tsupko}}},\ }\href {\doibase 10.1103/PhysRevD.106.064033} {\bibfield
  {journal} {\bibinfo  {journal} {\prd}\ }\textbf {\bibinfo {volume} {106}},\
  \bibinfo {pages} {064033} (\bibinfo {year} {2022})}\BibitemShut {NoStop}%
\bibitem [{\citenamefont {C\'ardenas-Avenda\~no}\ and\ \citenamefont
  {Lupsasca}(2023)}]{CardenasAvendano2023PRD108.064043}%
  \BibitemOpen
  \bibfield  {author} {\bibinfo {author} {\bibfnamefont {A.}~\bibnamefont
  {C\'ardenas-Avenda\~no}}\ and\ \bibinfo {author} {\bibfnamefont
  {A.}~\bibnamefont {Lupsasca}},\ }\href {\doibase 10.1103/PhysRevD.108.064043}
  {\bibfield  {journal} {\bibinfo  {journal} {\prd}\ }\textbf {\bibinfo
  {volume} {108}},\ \bibinfo {pages} {064043} (\bibinfo {year}
  {2023})}\BibitemShut {NoStop}%
\bibitem [{\citenamefont {Aratore}\ \emph {et~al.}(2024)\citenamefont
  {Aratore}, \citenamefont {Tsupko},\ and\ \citenamefont
  {Perlick}}]{Aratore2024PRD109.124057}%
  \BibitemOpen
  \bibfield  {author} {\bibinfo {author} {\bibfnamefont {F.}~\bibnamefont
  {Aratore}}, \bibinfo {author} {\bibfnamefont {O.~Y.}\ \bibnamefont {Tsupko}},
  \ and\ \bibinfo {author} {\bibfnamefont {V.}~\bibnamefont {Perlick}},\ }\href
  {\doibase 10.1103/PhysRevD.109.124057} {\bibfield  {journal} {\bibinfo
  {journal} {\prd}\ }\textbf {\bibinfo {volume} {109}},\ \bibinfo {pages}
  {124057} (\bibinfo {year} {2024})}\BibitemShut {NoStop}%
\bibitem [{\citenamefont {Jia}\ \emph {et~al.}(2024)\citenamefont {Jia},
  \citenamefont {Quataert}, \citenamefont {Lupsasca},\ and\ \citenamefont
  {Wong}}]{Jia2024PRD110.083044}%
  \BibitemOpen
  \bibfield  {author} {\bibinfo {author} {\bibfnamefont {H.}~\bibnamefont
  {Jia}}, \bibinfo {author} {\bibfnamefont {E.}~\bibnamefont {Quataert}},
  \bibinfo {author} {\bibfnamefont {A.}~\bibnamefont {Lupsasca}}, \ and\
  \bibinfo {author} {\bibfnamefont {G.~N.}\ \bibnamefont {Wong}},\ }\href
  {\doibase 10.1103/PhysRevD.110.083044} {\bibfield  {journal} {\bibinfo
  {journal} {\prd}\ }\textbf {\bibinfo {volume} {110}},\ \bibinfo {pages}
  {083044} (\bibinfo {year} {2024})}\BibitemShut {NoStop}%
\bibitem [{\citenamefont {{Falcke}}\ \emph {et~al.}(2000)\citenamefont
  {{Falcke}}, \citenamefont {{Melia}},\ and\ \citenamefont
  {{Agol}}}]{Falcke2000ApJ528.L13}%
  \BibitemOpen
  \bibfield  {author} {\bibinfo {author} {\bibfnamefont {H.}~\bibnamefont
  {{Falcke}}}, \bibinfo {author} {\bibfnamefont {F.}~\bibnamefont {{Melia}}}, \
  and\ \bibinfo {author} {\bibfnamefont {E.}~\bibnamefont {{Agol}}},\ }\href
  {\doibase 10.1086/312423} {\bibfield  {journal} {\bibinfo  {journal} {\apjl}\
  }\textbf {\bibinfo {volume} {528}},\ \bibinfo {pages} {L13} (\bibinfo {year}
  {2000})}\BibitemShut {NoStop}%
\bibitem [{\citenamefont {{Cunha}}\ and\ \citenamefont
  {{Herdeiro}}(2018)}]{Cunha2018GRG50.42}%
  \BibitemOpen
  \bibfield  {author} {\bibinfo {author} {\bibfnamefont {P.~V.~P.}\
  \bibnamefont {{Cunha}}}\ and\ \bibinfo {author} {\bibfnamefont {C.~A.~R.}\
  \bibnamefont {{Herdeiro}}},\ }\href {\doibase 10.1007/s10714-018-2361-9}
  {\bibfield  {journal} {\bibinfo  {journal} {\grg}\ }\textbf {\bibinfo
  {volume} {50}},\ \bibinfo {eid} {42} (\bibinfo {year} {2018})}\BibitemShut
  {NoStop}%
\bibitem [{\citenamefont {{Perlick}}\ and\ \citenamefont
  {{Tsupko}}(2022)}]{Perlick2022PR947.1}%
  \BibitemOpen
  \bibfield  {author} {\bibinfo {author} {\bibfnamefont {V.}~\bibnamefont
  {{Perlick}}}\ and\ \bibinfo {author} {\bibfnamefont {O.~Y.}\ \bibnamefont
  {{Tsupko}}},\ }\href {\doibase 10.1016/j.physrep.2021.10.004} {\bibfield
  {journal} {\bibinfo  {journal} {\physrep}\ }\textbf {\bibinfo {volume}
  {947}},\ \bibinfo {pages} {1} (\bibinfo {year} {2022})}\BibitemShut {NoStop}%
\bibitem [{\citenamefont {{Bozza}}(2002)}]{Bozza2002PRD66.103001}%
  \BibitemOpen
  \bibfield  {author} {\bibinfo {author} {\bibfnamefont {V.}~\bibnamefont
  {{Bozza}}},\ }\href {\doibase 10.1103/PhysRevD.66.103001} {\bibfield
  {journal} {\bibinfo  {journal} {\prd}\ }\textbf {\bibinfo {volume} {66}},\
  \bibinfo {eid} {103001} (\bibinfo {year} {2002})}\BibitemShut {NoStop}%
\bibitem [{\citenamefont {{Patil}}\ \emph {et~al.}(2017)\citenamefont
  {{Patil}}, \citenamefont {{Mishra}},\ and\ \citenamefont
  {{Narasimha}}}]{Patil2017PRD95.024026}%
  \BibitemOpen
  \bibfield  {author} {\bibinfo {author} {\bibfnamefont {M.}~\bibnamefont
  {{Patil}}}, \bibinfo {author} {\bibfnamefont {P.}~\bibnamefont {{Mishra}}}, \
  and\ \bibinfo {author} {\bibfnamefont {D.}~\bibnamefont {{Narasimha}}},\
  }\href {\doibase 10.1103/PhysRevD.95.024026} {\bibfield  {journal} {\bibinfo
  {journal} {\prd}\ }\textbf {\bibinfo {volume} {95}},\ \bibinfo {eid} {024026}
  (\bibinfo {year} {2017})}\BibitemShut {NoStop}%
\bibitem [{\citenamefont {Gyulchev}\ \emph {et~al.}(2021)\citenamefont
  {Gyulchev}, \citenamefont {Nedkova}, \citenamefont {Vetsov},\ and\
  \citenamefont {Yazadjiev}}]{Gyulchev2021EPJC81.885}%
  \BibitemOpen
  \bibfield  {author} {\bibinfo {author} {\bibfnamefont {G.}~\bibnamefont
  {Gyulchev}}, \bibinfo {author} {\bibfnamefont {P.}~\bibnamefont {Nedkova}},
  \bibinfo {author} {\bibfnamefont {T.}~\bibnamefont {Vetsov}}, \ and\ \bibinfo
  {author} {\bibfnamefont {S.}~\bibnamefont {Yazadjiev}},\ }\href {\doibase
  10.1140/epjc/s10052-021-09624-5} {\bibfield  {journal} {\bibinfo  {journal}
  {\epjc}\ }\textbf {\bibinfo {volume} {81}},\ \bibinfo {pages} {885} (\bibinfo
  {year} {2021})}\BibitemShut {NoStop}%
\bibitem [{\citenamefont {{Zhu}}\ and\ \citenamefont
  {{Xie}}(2020)}]{Zhu2020EPJC80.444}%
  \BibitemOpen
  \bibfield  {author} {\bibinfo {author} {\bibfnamefont {X.-Y.}\ \bibnamefont
  {{Zhu}}}\ and\ \bibinfo {author} {\bibfnamefont {Y.}~\bibnamefont {{Xie}}},\
  }\href {\doibase 10.1140/epjc/s10052-020-8021-8} {\bibfield  {journal}
  {\bibinfo  {journal} {\epjc}\ }\textbf {\bibinfo {volume} {80}},\ \bibinfo
  {eid} {444} (\bibinfo {year} {2020})}\BibitemShut {NoStop}%
\bibitem [{\citenamefont {Igata}\ \emph {et~al.}(2025)\citenamefont {Igata},
  \citenamefont {Omamiuda},\ and\ \citenamefont
  {Takamori}}]{Igata2025PRD111.084062}%
  \BibitemOpen
  \bibfield  {author} {\bibinfo {author} {\bibfnamefont {T.}~\bibnamefont
  {Igata}}, \bibinfo {author} {\bibfnamefont {M.}~\bibnamefont {Omamiuda}}, \
  and\ \bibinfo {author} {\bibfnamefont {Y.}~\bibnamefont {Takamori}},\ }\href
  {\doibase 10.1103/PhysRevD.111.084062} {\bibfield  {journal} {\bibinfo
  {journal} {\prd}\ }\textbf {\bibinfo {volume} {111}},\ \bibinfo {pages}
  {084062} (\bibinfo {year} {2025})}\BibitemShut {NoStop}%
\bibitem [{\citenamefont {Tamm}\ and\ \citenamefont
  {Rosa}(2024)}]{Tamm2024PRD109.044062}%
  \BibitemOpen
  \bibfield  {author} {\bibinfo {author} {\bibfnamefont {H.~L.}\ \bibnamefont
  {Tamm}}\ and\ \bibinfo {author} {\bibfnamefont {J.~a.~L.}\ \bibnamefont
  {Rosa}},\ }\href {\doibase 10.1103/PhysRevD.109.044062} {\bibfield  {journal}
  {\bibinfo  {journal} {\prd}\ }\textbf {\bibinfo {volume} {109}},\ \bibinfo
  {pages} {044062} (\bibinfo {year} {2024})}\BibitemShut {NoStop}%
\bibitem [{\citenamefont {Olmo}\ \emph {et~al.}(2022)\citenamefont {Olmo},
  \citenamefont {Rubiera-Garcia},\ and\ \citenamefont
  {G\'omez}}]{Olmo2021PLB829.137045}%
  \BibitemOpen
  \bibfield  {author} {\bibinfo {author} {\bibfnamefont {G.~J.}\ \bibnamefont
  {Olmo}}, \bibinfo {author} {\bibfnamefont {D.}~\bibnamefont
  {Rubiera-Garcia}}, \ and\ \bibinfo {author} {\bibfnamefont {D.~S.-C.}\
  \bibnamefont {G\'omez}},\ }\href {\doibase 10.1016/j.physletb.2022.137045}
  {\bibfield  {journal} {\bibinfo  {journal} {\plb}\ }\textbf {\bibinfo
  {volume} {829}},\ \bibinfo {pages} {137045} (\bibinfo {year}
  {2022})}\BibitemShut {NoStop}%
\bibitem [{\citenamefont {{Rosa}}(2023)}]{Rosa2023PRD107.084048}%
  \BibitemOpen
  \bibfield  {author} {\bibinfo {author} {\bibfnamefont {J.~L.}\ \bibnamefont
  {{Rosa}}},\ }\href {\doibase 10.1103/PhysRevD.107.084048} {\bibfield
  {journal} {\bibinfo  {journal} {\prd}\ }\textbf {\bibinfo {volume} {107}},\
  \bibinfo {pages} {084048} (\bibinfo {year} {2023})}\BibitemShut {NoStop}%
\bibitem [{\citenamefont {{Weinberg}}(1972)}]{Weinberg1972Book}%
  \BibitemOpen
  \bibfield  {author} {\bibinfo {author} {\bibfnamefont {S.}~\bibnamefont
  {{Weinberg}}},\ }\href@noop {} {\emph {\bibinfo {title} {Gravitation and
  Cosmology: Principles and Applications of the General Theory of
  Relativity}}}\ (\bibinfo  {publisher} {{Wiley}},\ \bibinfo {address} {{New
  York}},\ \bibinfo {year} {1972})\BibitemShut {NoStop}%
\bibitem [{\citenamefont {{Virbhadra}}\ \emph {et~al.}(1998)\citenamefont
  {{Virbhadra}}, \citenamefont {{Narasimha}},\ and\ \citenamefont
  {{Chitre}}}]{Virbhadra1998AA337.1}%
  \BibitemOpen
  \bibfield  {author} {\bibinfo {author} {\bibfnamefont {K.~S.}\ \bibnamefont
  {{Virbhadra}}}, \bibinfo {author} {\bibfnamefont {D.}~\bibnamefont
  {{Narasimha}}}, \ and\ \bibinfo {author} {\bibfnamefont {S.~M.}\ \bibnamefont
  {{Chitre}}},\ }\href@noop {} {\bibfield  {journal} {\bibinfo  {journal}
  {\aap}\ }\textbf {\bibinfo {volume} {337}},\ \bibinfo {pages} {1} (\bibinfo
  {year} {1998})}\BibitemShut {NoStop}%
\bibitem [{\citenamefont {{Cunha}}\ \emph {et~al.}(2017)\citenamefont
  {{Cunha}}, \citenamefont {{Berti}},\ and\ \citenamefont
  {{Herdeiro}}}]{Cunha2017PRL119.251102}%
  \BibitemOpen
  \bibfield  {author} {\bibinfo {author} {\bibfnamefont {P.~V.~P.}\
  \bibnamefont {{Cunha}}}, \bibinfo {author} {\bibfnamefont {E.}~\bibnamefont
  {{Berti}}}, \ and\ \bibinfo {author} {\bibfnamefont {C.~A.~R.}\ \bibnamefont
  {{Herdeiro}}},\ }\href {\doibase 10.1103/PhysRevLett.119.251102} {\bibfield
  {journal} {\bibinfo  {journal} {\prl}\ }\textbf {\bibinfo {volume} {119}},\
  \bibinfo {eid} {251102} (\bibinfo {year} {2017})}\BibitemShut {NoStop}%
\bibitem [{\citenamefont {Guo}\ and\ \citenamefont
  {Gao}(2021)}]{Guo2020PRD103.104031}%
  \BibitemOpen
  \bibfield  {author} {\bibinfo {author} {\bibfnamefont {M.}~\bibnamefont
  {Guo}}\ and\ \bibinfo {author} {\bibfnamefont {S.}~\bibnamefont {Gao}},\
  }\href {\doibase 10.1103/PhysRevD.103.104031} {\bibfield  {journal} {\bibinfo
   {journal} {\prd}\ }\textbf {\bibinfo {volume} {103}},\ \bibinfo {pages}
  {104031} (\bibinfo {year} {2021})}\BibitemShut {NoStop}%
\bibitem [{\citenamefont {Cunha}\ \emph {et~al.}(2016)\citenamefont {Cunha},
  \citenamefont {Grover}, \citenamefont {Herdeiro}, \citenamefont {Radu},
  \citenamefont {Runarsson},\ and\ \citenamefont
  {Wittig}}]{Cunha2016PRD94.104023}%
  \BibitemOpen
  \bibfield  {author} {\bibinfo {author} {\bibfnamefont {P.~V.~P.}\
  \bibnamefont {Cunha}}, \bibinfo {author} {\bibfnamefont {J.}~\bibnamefont
  {Grover}}, \bibinfo {author} {\bibfnamefont {C.}~\bibnamefont {Herdeiro}},
  \bibinfo {author} {\bibfnamefont {E.}~\bibnamefont {Radu}}, \bibinfo {author}
  {\bibfnamefont {H.}~\bibnamefont {Runarsson}}, \ and\ \bibinfo {author}
  {\bibfnamefont {A.}~\bibnamefont {Wittig}},\ }\href {\doibase
  10.1103/PhysRevD.94.104023} {\bibfield  {journal} {\bibinfo  {journal}
  {\prd}\ }\textbf {\bibinfo {volume} {94}},\ \bibinfo {pages} {104023}
  (\bibinfo {year} {2016})}\BibitemShut {NoStop}%
\bibitem [{\citenamefont {{Luminet}}(1979)}]{Luminet1979AA75.228}%
  \BibitemOpen
  \bibfield  {author} {\bibinfo {author} {\bibfnamefont {J.-P.}\ \bibnamefont
  {{Luminet}}},\ }\href@noop {} {\bibfield  {journal} {\bibinfo  {journal}
  {\aap}\ }\textbf {\bibinfo {volume} {75}},\ \bibinfo {pages} {228} (\bibinfo
  {year} {1979})}\BibitemShut {NoStop}%
\bibitem [{\citenamefont {{Thompson}}\ \emph {et~al.}(2017)\citenamefont
  {{Thompson}}, \citenamefont {{Moran}},\ and\ \citenamefont
  {{Swenson}}}]{Thompson2017Book}%
  \BibitemOpen
  \bibfield  {author} {\bibinfo {author} {\bibfnamefont {A.~R.}\ \bibnamefont
  {{Thompson}}}, \bibinfo {author} {\bibfnamefont {J.~M.}\ \bibnamefont
  {{Moran}}}, \ and\ \bibinfo {author} {\bibfnamefont {G.~W.}\ \bibnamefont
  {{Swenson}}},\ }\href {\doibase 10.1007/978-3-319-44431-4} {\emph {\bibinfo
  {title} {{Interferometry and Synthesis in Radio Astronomy}}}}\ (\bibinfo
  {publisher} {Springer, Cham},\ \bibinfo {year} {2017})\BibitemShut {NoStop}%
\bibitem [{\citenamefont {{Gralla}}(2020)}]{Gralla2020PRD102.044017}%
  \BibitemOpen
  \bibfield  {author} {\bibinfo {author} {\bibfnamefont {S.~E.}\ \bibnamefont
  {{Gralla}}},\ }\href {\doibase 10.1103/PhysRevD.102.044017} {\bibfield
  {journal} {\bibinfo  {journal} {\prd}\ }\textbf {\bibinfo {volume} {102}},\
  \bibinfo {eid} {044017} (\bibinfo {year} {2020})}\BibitemShut {NoStop}%
\bibitem [{\citenamefont {Gralla}\ and\ \citenamefont
  {Lupsasca}(2020{\natexlab{a}})}]{Gralla2020PRD102.124003}%
  \BibitemOpen
  \bibfield  {author} {\bibinfo {author} {\bibfnamefont {S.~E.}\ \bibnamefont
  {Gralla}}\ and\ \bibinfo {author} {\bibfnamefont {A.}~\bibnamefont
  {Lupsasca}},\ }\href {\doibase 10.1103/PhysRevD.102.124003} {\bibfield
  {journal} {\bibinfo  {journal} {\prd}\ }\textbf {\bibinfo {volume} {102}},\
  \bibinfo {pages} {124003} (\bibinfo {year} {2020}{\natexlab{a}})}\BibitemShut
  {NoStop}%
\bibitem [{\citenamefont {{Stephani}}\ \emph {et~al.}(2009)\citenamefont
  {{Stephani}}, \citenamefont {{Kramer}}, \citenamefont {{MacCallum}},
  \citenamefont {{Hoenselaers}},\ and\ \citenamefont
  {{Herlt}}}]{Stephani2009book}%
  \BibitemOpen
  \bibfield  {author} {\bibinfo {author} {\bibfnamefont {H.}~\bibnamefont
  {{Stephani}}}, \bibinfo {author} {\bibfnamefont {D.}~\bibnamefont
  {{Kramer}}}, \bibinfo {author} {\bibfnamefont {M.}~\bibnamefont
  {{MacCallum}}}, \bibinfo {author} {\bibfnamefont {C.}~\bibnamefont
  {{Hoenselaers}}}, \ and\ \bibinfo {author} {\bibfnamefont {E.}~\bibnamefont
  {{Herlt}}},\ }\href {\doibase 10.1017/CBO9780511535185} {\emph {\bibinfo
  {title} {{Exact Solutions of Einstein's Field Equations}}}}\ (\bibinfo {year}
  {2009})\BibitemShut {NoStop}%
\bibitem [{\citenamefont {{Buchdahl}}(1959)}]{Buchdahl1959PR116.1027}%
  \BibitemOpen
  \bibfield  {author} {\bibinfo {author} {\bibfnamefont {H.~A.}\ \bibnamefont
  {{Buchdahl}}},\ }\href {\doibase 10.1103/PhysRev.116.1027} {\bibfield
  {journal} {\bibinfo  {journal} {\pr}\ }\textbf {\bibinfo {volume} {116}},\
  \bibinfo {pages} {1027} (\bibinfo {year} {1959})}\BibitemShut {NoStop}%
\bibitem [{\citenamefont {Gabbanelli}\ \emph {et~al.}(2019)\citenamefont
  {Gabbanelli}, \citenamefont {Ovalle}, \citenamefont {Sotomayor},
  \citenamefont {Stuchlik},\ and\ \citenamefont
  {Casadio}}]{Gabbanelli2019EPJC79.486}%
  \BibitemOpen
  \bibfield  {author} {\bibinfo {author} {\bibfnamefont {L.}~\bibnamefont
  {Gabbanelli}}, \bibinfo {author} {\bibfnamefont {J.}~\bibnamefont {Ovalle}},
  \bibinfo {author} {\bibfnamefont {A.}~\bibnamefont {Sotomayor}}, \bibinfo
  {author} {\bibfnamefont {Z.}~\bibnamefont {Stuchlik}}, \ and\ \bibinfo
  {author} {\bibfnamefont {R.}~\bibnamefont {Casadio}},\ }\href {\doibase
  10.1140/epjc/s10052-019-7022-y} {\bibfield  {journal} {\bibinfo  {journal}
  {\epjc}\ }\textbf {\bibinfo {volume} {79}},\ \bibinfo {pages} {486} (\bibinfo
  {year} {2019})}\BibitemShut {NoStop}%
\bibitem [{\citenamefont {{Ovalle}}\ \emph {et~al.}(2019)\citenamefont
  {{Ovalle}}, \citenamefont {{Posada}},\ and\ \citenamefont
  {{Stuchl{\'\i}k}}}]{Ovalle2019CQG36.205010}%
  \BibitemOpen
  \bibfield  {author} {\bibinfo {author} {\bibfnamefont {J.}~\bibnamefont
  {{Ovalle}}}, \bibinfo {author} {\bibfnamefont {C.}~\bibnamefont {{Posada}}},
  \ and\ \bibinfo {author} {\bibfnamefont {Z.}~\bibnamefont
  {{Stuchl{\'\i}k}}},\ }\href {\doibase 10.1088/1361-6382/ab4461} {\bibfield
  {journal} {\bibinfo  {journal} {\cqg}\ }\textbf {\bibinfo {volume} {36}},\
  \bibinfo {eid} {205010} (\bibinfo {year} {2019})}\BibitemShut {NoStop}%
\bibitem [{\citenamefont {{Beltracchi}}\ and\ \citenamefont
  {{Gondolo}}(2019)}]{Beltracchi2019PRD99.084021}%
  \BibitemOpen
  \bibfield  {author} {\bibinfo {author} {\bibfnamefont {P.}~\bibnamefont
  {{Beltracchi}}}\ and\ \bibinfo {author} {\bibfnamefont {P.}~\bibnamefont
  {{Gondolo}}},\ }\href {\doibase 10.1103/PhysRevD.99.084021} {\bibfield
  {journal} {\bibinfo  {journal} {\prd}\ }\textbf {\bibinfo {volume} {99}},\
  \bibinfo {pages} {084021} (\bibinfo {year} {2019})}\BibitemShut {NoStop}%
\bibitem [{\citenamefont {{Mazur}}\ and\ \citenamefont
  {{Mottola}}(2023)}]{Mazur2023Universe9.88}%
  \BibitemOpen
  \bibfield  {author} {\bibinfo {author} {\bibfnamefont {P.~O.}\ \bibnamefont
  {{Mazur}}}\ and\ \bibinfo {author} {\bibfnamefont {E.}~\bibnamefont
  {{Mottola}}},\ }\href {\doibase 10.3390/universe9020088} {\bibfield
  {journal} {\bibinfo  {journal} {Universe}\ }\textbf {\bibinfo {volume} {9}},\
  \bibinfo {pages} {88} (\bibinfo {year} {2023})}\BibitemShut {NoStop}%
\bibitem [{\citenamefont {{Mazur}}\ and\ \citenamefont
  {{Mottola}}(2004)}]{Mazur2004PNAS101.9545}%
  \BibitemOpen
  \bibfield  {author} {\bibinfo {author} {\bibfnamefont {P.~O.}\ \bibnamefont
  {{Mazur}}}\ and\ \bibinfo {author} {\bibfnamefont {E.}~\bibnamefont
  {{Mottola}}},\ }\href {\doibase 10.1073/pnas.0402717101} {\bibfield
  {journal} {\bibinfo  {journal} {\pnas}\ }\textbf {\bibinfo {volume} {101}},\
  \bibinfo {pages} {9545} (\bibinfo {year} {2004})}\BibitemShut {NoStop}%
\bibitem [{\citenamefont {{Chirenti}}\ \emph {et~al.}(2020)\citenamefont
  {{Chirenti}}, \citenamefont {{Posada}},\ and\ \citenamefont
  {{Guedes}}}]{Chirenti2020CQG37.195017}%
  \BibitemOpen
  \bibfield  {author} {\bibinfo {author} {\bibfnamefont {C.}~\bibnamefont
  {{Chirenti}}}, \bibinfo {author} {\bibfnamefont {C.}~\bibnamefont
  {{Posada}}}, \ and\ \bibinfo {author} {\bibfnamefont {V.}~\bibnamefont
  {{Guedes}}},\ }\href {\doibase 10.1088/1361-6382/abb07a} {\bibfield
  {journal} {\bibinfo  {journal} {\cqg}\ }\textbf {\bibinfo {volume} {37}},\
  \bibinfo {pages} {195017} (\bibinfo {year} {2020})}\BibitemShut {NoStop}%
\bibitem [{\citenamefont {{Konoplya}}\ \emph {et~al.}(2019)\citenamefont
  {{Konoplya}}, \citenamefont {{Posada}}, \citenamefont {{Stuchl{\'\i}k}},\
  and\ \citenamefont {{Zhidenko}}}]{Konoplya2019PRD100.044027}%
  \BibitemOpen
  \bibfield  {author} {\bibinfo {author} {\bibfnamefont {R.~A.}\ \bibnamefont
  {{Konoplya}}}, \bibinfo {author} {\bibfnamefont {C.}~\bibnamefont
  {{Posada}}}, \bibinfo {author} {\bibfnamefont {Z.}~\bibnamefont
  {{Stuchl{\'\i}k}}}, \ and\ \bibinfo {author} {\bibfnamefont {A.}~\bibnamefont
  {{Zhidenko}}},\ }\href {\doibase 10.1103/PhysRevD.100.044027} {\bibfield
  {journal} {\bibinfo  {journal} {\prd}\ }\textbf {\bibinfo {volume} {100}},\
  \bibinfo {pages} {044027} (\bibinfo {year} {2019})}\BibitemShut {NoStop}%
\bibitem [{\citenamefont {{Sakai}}\ \emph {et~al.}(2014)\citenamefont
  {{Sakai}}, \citenamefont {{Saida}},\ and\ \citenamefont
  {{Tamaki}}}]{Sakai2014PRD90.104013}%
  \BibitemOpen
  \bibfield  {author} {\bibinfo {author} {\bibfnamefont {N.}~\bibnamefont
  {{Sakai}}}, \bibinfo {author} {\bibfnamefont {H.}~\bibnamefont {{Saida}}}, \
  and\ \bibinfo {author} {\bibfnamefont {T.}~\bibnamefont {{Tamaki}}},\ }\href
  {\doibase 10.1103/PhysRevD.90.104013} {\bibfield  {journal} {\bibinfo
  {journal} {\prd}\ }\textbf {\bibinfo {volume} {90}},\ \bibinfo {eid} {104013}
  (\bibinfo {year} {2014})}\BibitemShut {NoStop}%
\bibitem [{\citenamefont {Roelofs}\ \emph {et~al.}(2023)\citenamefont {Roelofs}
  \emph {et~al.}}]{Roelofs2022Gal11.12}%
  \BibitemOpen
  \bibfield  {author} {\bibinfo {author} {\bibfnamefont {F.}~\bibnamefont
  {Roelofs}} \emph {et~al.},\ }\href {\doibase 10.3390/galaxies11010012}
  {\bibfield  {journal} {\bibinfo  {journal} {Galaxies}\ }\textbf {\bibinfo
  {volume} {11}},\ \bibinfo {pages} {12} (\bibinfo {year} {2023})}\BibitemShut
  {NoStop}%
\bibitem [{\citenamefont {{Psaltis \textit{et al.} (Event Horizon Telescope
  Collaboration)}}(2020)}]{EHTC2020PRL125.141104}%
  \BibitemOpen
  \bibfield  {author} {\bibinfo {author} {\bibfnamefont {D.}~\bibnamefont
  {{Psaltis \textit{et al.} (Event Horizon Telescope Collaboration)}}},\ }\href
  {\doibase 10.1103/PhysRevLett.125.141104} {\bibfield  {journal} {\bibinfo
  {journal} {\prl}\ }\textbf {\bibinfo {volume} {125}},\ \bibinfo {pages}
  {141104} (\bibinfo {year} {2020})}\BibitemShut {NoStop}%
\bibitem [{\citenamefont {Zhang}\ \emph {et~al.}(2025)\citenamefont {Zhang},
  \citenamefont {Hou}, \citenamefont {Guo}, \citenamefont {Mizuno},\ and\
  \citenamefont {Chen}}]{Zhang2025arXiv2503.17200}%
  \BibitemOpen
  \bibfield  {author} {\bibinfo {author} {\bibfnamefont {Z.}~\bibnamefont
  {Zhang}}, \bibinfo {author} {\bibfnamefont {Y.}~\bibnamefont {Hou}}, \bibinfo
  {author} {\bibfnamefont {M.}~\bibnamefont {Guo}}, \bibinfo {author}
  {\bibfnamefont {Y.}~\bibnamefont {Mizuno}}, \ and\ \bibinfo {author}
  {\bibfnamefont {B.}~\bibnamefont {Chen}},\ }\href@noop {} {\  (\bibinfo
  {year} {2025})},\ \Eprint {http://arxiv.org/abs/2503.17200} {arXiv:2503.17200
  [astro-ph.HE]} \BibitemShut {NoStop}%
\bibitem [{\citenamefont {Zhu}(2025)}]{Zhu2025arXiv2503.22343}%
  \BibitemOpen
  \bibfield  {author} {\bibinfo {author} {\bibfnamefont {Q.-H.}\ \bibnamefont
  {Zhu}},\ }\href@noop {} {\  (\bibinfo {year} {2025})},\ \Eprint
  {http://arxiv.org/abs/2503.22343} {arXiv:2503.22343 [astro-ph.HE]}
  \BibitemShut {NoStop}%
\bibitem [{\citenamefont {{Posada}}(2017)}]{Posada2017MNRAS468.2128}%
  \BibitemOpen
  \bibfield  {author} {\bibinfo {author} {\bibfnamefont {C.}~\bibnamefont
  {{Posada}}},\ }\href {\doibase 10.1093/mnras/stx523} {\bibfield  {journal}
  {\bibinfo  {journal} {\mnras}\ }\textbf {\bibinfo {volume} {468}},\ \bibinfo
  {pages} {2128} (\bibinfo {year} {2017})}\BibitemShut {NoStop}%
\bibitem [{\citenamefont {Hernandez-Pastora}\ and\ \citenamefont
  {Herrera}(2017)}]{Hernandez-Pastora2017PRD95.024003}%
  \BibitemOpen
  \bibfield  {author} {\bibinfo {author} {\bibfnamefont {J.~L.}\ \bibnamefont
  {Hernandez-Pastora}}\ and\ \bibinfo {author} {\bibfnamefont {L.}~\bibnamefont
  {Herrera}},\ }\href {\doibase 10.1103/PhysRevD.95.024003} {\bibfield
  {journal} {\bibinfo  {journal} {\prd}\ }\textbf {\bibinfo {volume} {95}},\
  \bibinfo {pages} {024003} (\bibinfo {year} {2017})}\BibitemShut {NoStop}%
\bibitem [{\citenamefont {{Ravi}}\ and\ \citenamefont
  {{Banerjee}}(2018)}]{Ravi2018NA64.31}%
  \BibitemOpen
  \bibfield  {author} {\bibinfo {author} {\bibfnamefont {A.~P.}\ \bibnamefont
  {{Ravi}}}\ and\ \bibinfo {author} {\bibfnamefont {N.}~\bibnamefont
  {{Banerjee}}},\ }\href {\doibase 10.1016/j.newast.2018.04.003} {\bibfield
  {journal} {\bibinfo  {journal} {\na}\ }\textbf {\bibinfo {volume} {64}},\
  \bibinfo {pages} {31} (\bibinfo {year} {2018})}\BibitemShut {NoStop}%
\bibitem [{\citenamefont {{Kim}}\ \emph {et~al.}(2020)\citenamefont {{Kim}},
  \citenamefont {{Lee}}, \citenamefont {{Lee}},\ and\ \citenamefont
  {{Lee}}}]{Kim2020PRD101.064067}%
  \BibitemOpen
  \bibfield  {author} {\bibinfo {author} {\bibfnamefont {H.}~\bibnamefont
  {{Kim}}}, \bibinfo {author} {\bibfnamefont {B.}~\bibnamefont {{Lee}}},
  \bibinfo {author} {\bibfnamefont {W.}~\bibnamefont {{Lee}}}, \ and\ \bibinfo
  {author} {\bibfnamefont {Y.}~\bibnamefont {{Lee}}},\ }\href {\doibase
  10.1103/PhysRevD.101.064067} {\bibfield  {journal} {\bibinfo  {journal}
  {\prd}\ }\textbf {\bibinfo {volume} {101}},\ \bibinfo {pages} {064067}
  (\bibinfo {year} {2020})}\BibitemShut {NoStop}%
\bibitem [{\citenamefont {Viaggiu}(2023)}]{Viaggiu2023IJMPD32.2350008}%
  \BibitemOpen
  \bibfield  {author} {\bibinfo {author} {\bibfnamefont {S.}~\bibnamefont
  {Viaggiu}},\ }\href {\doibase 10.1142/S0218271823500086} {\bibfield
  {journal} {\bibinfo  {journal} {\ijmpd}\ }\textbf {\bibinfo {volume} {32}},\
  \bibinfo {pages} {2350008} (\bibinfo {year} {2023})}\BibitemShut {NoStop}%
\bibitem [{\citenamefont {Gralla}\ and\ \citenamefont
  {Lupsasca}(2020{\natexlab{b}})}]{Gralla2020PRD101.044031}%
  \BibitemOpen
  \bibfield  {author} {\bibinfo {author} {\bibfnamefont {S.~E.}\ \bibnamefont
  {Gralla}}\ and\ \bibinfo {author} {\bibfnamefont {A.}~\bibnamefont
  {Lupsasca}},\ }\href {\doibase 10.1103/PhysRevD.101.044031} {\bibfield
  {journal} {\bibinfo  {journal} {\prd}\ }\textbf {\bibinfo {volume} {101}},\
  \bibinfo {pages} {044031} (\bibinfo {year} {2020}{\natexlab{b}})}\BibitemShut
  {NoStop}%
\bibitem [{\citenamefont {Zhou}\ \emph {et~al.}(2025)\citenamefont {Zhou},
  \citenamefont {Zhong}, \citenamefont {Chen},\ and\ \citenamefont
  {Cardoso}}]{Zhou2025PRD111.064075}%
  \BibitemOpen
  \bibfield  {author} {\bibinfo {author} {\bibfnamefont {L.}~\bibnamefont
  {Zhou}}, \bibinfo {author} {\bibfnamefont {Z.}~\bibnamefont {Zhong}},
  \bibinfo {author} {\bibfnamefont {Y.}~\bibnamefont {Chen}}, \ and\ \bibinfo
  {author} {\bibfnamefont {V.}~\bibnamefont {Cardoso}},\ }\href {\doibase
  10.1103/PhysRevD.111.064075} {\bibfield  {journal} {\bibinfo  {journal}
  {\prd}\ }\textbf {\bibinfo {volume} {111}},\ \bibinfo {pages} {064075}
  (\bibinfo {year} {2025})}\BibitemShut {NoStop}%
\end{thebibliography}%

\end{document}